%% file: Hagh.tex
\documentclass[vecphys]{svmult}

\usepackage{makeidx}         
\usepackage{graphicx}        
\usepackage{multicol}        
\usepackage[bottom]{footmisc}
\makeindex             
\begin{document}

\title*{Planetary Dynamics and Habitable Planet Formation In Binary Star Systems}
\titlerunning{Habitable Planets in Binaries}
\author{Nader Haghighipour\inst{1},
Rudolf Dvorak\inst{2}, \and  Elke Pilat-Lohinger\inst{2}}
\institute{Institute for Astronomy and NASA Astrobiology Institute,
University of Hawaii-Manoa,
(\texttt{nader@ifa.hawaii.edu})
\and Institute for Astronomy, University of Vienna,
\texttt{(dvorak@astro.univie.ac.at, lohinger@astro.univie.ac.at)}}
\authorrunning{Haghighipour, Dvorak, Pilat-Lohinger}

\maketitle

\section{Introduction}
\label{sec:1}

How our planet was formed, how life came about, and whether life exists elsewhere 
in the universe are among some of the long-standing questions in human history. The latter, 
which has been the main drive behind many decades of searching for planets outside 
the solar system, is one of the most outstanding problems in planetary science and 
astrobiology. Although no Earth-like planet has yet been found, the success of 
observational techniques in identifying now more than 350  extrasolar planets has 
greatly contributed to addressing this question, and has extended the concept of
habitability to billions of miles beyond the boundaries of our solar system. It is now 
certain that our planetary system is not unique and many terrestrial-size planets may 
exist throughout the universe. 

The orbital and physical diversity of the currently known extrasolar planets play a 
crucial role in their habitability. In general, whether a planet can be habitable depends 
on its physical and dynamical properties, and the luminosity of its host star. The notion 
of habitability is normally defined based on the life as we know it, and uses the physical 
and orbital characteristics of Earth as an example of a habitable planet. In other words, 
a planet is habitable if it is Earth-like so it can develop and sustain Earthly life. This 
definition of habitability requires that a potentially habitable planet to maintain liquid water 
on its surface and in its atmosphere. The planet's capability in maintaining water is 
determined by its size and orbital motion, the luminosity of the central star, and the
distribution of water in the circumstellar material from which the planet was formed.

How extrasolar habitable planets are formed is a widely addressed question that is 
still unresolved. While models of planetary accretion in the inner solar system present 
pathways (although in some cases incomplete) toward the formation of planets such 
as Earth and Venus, the orbital diversity of extrasolar planets present strong challenges 
to the applicability of these models to other planetary environments. For instance,
systems with close-in giant planets may require massive protoplanetary disks to 
ensure that while planetesimals and protoplanets are scattered as giant planets migrate, 
terrestrial bodies can form and be stable. Systems with multiple planets also present a great 
challenge to terrestrial planet formation since the orbital architectures of such systems 
may limit the regions of the stability of smaller objects. 

In a system with two stars, the situation is even more complicated. The interaction 
between one star and the protoplanetary disk around the other may inhibit planet 
formation by truncating the disk and removing circumstellar material 
\cite{Artymowicz94}. This interaction may also  prevent the growth of km-size 
planetesimals to larger objects by increasing the relative velocities of these bodies
and causing their collisions to result in fragmentation. 
Despite such difficulties, planets have, however, been detected in binary star systems 
(see \ref{tab:1}) and observers have been able to identify three 
moderately close ($< 20$ AU) binaries, namely $\gamma$ Cephei \cite{Hat03}, 
GL 86 \cite{Eggenberger01} , and HD 41004 \cite{Zucker04}, whose primary stars 
are hosts to Jupiter-like planets. 

The detection of planets in binary star systems is not a surprise. There is much 
observational evidence that indicates the most common outcome of the star formation 
process is a binary system \cite{Math94,White01}. Also, as shown by Prato \& Weinberger
in chapter 1, there is substantial evidence for the existence of potentially planet-forming circumstellar disks in multiple star systems 
\cite{Math94,Akeson98,Rodriguez98,White99,Silbert00,Math00,Trilling07}. 
These all point to the fact that planet formation in binaries is robust and many of these 
systems may harbor additional giant planets and/or terrestrial-size objects.
This chapter is devoted to study the latter.

Whether binaries can harbor potentially habitable planets depends on several 
factors including the physical properties and the orbital characteristics of the binary 
system. While the former determines the location of the habitable zone (HZ), the 
latter affects the dynamics of the material from which terrestrial planets are formed 
(i.e., planetesimals and planetary embryos), and drives the final architecture of the 
planets assembly. In order for a habitable planet to form in a binary star system,
these two factors have to work in harmony. That is, the orbital dynamics of the two 
stars and their interactions with the planet-forming material have to allow terrestrial 
planet formation in the habitable zone, and ensure that the orbit of a potentially 
habitable planet will be stable for long times. We organize this chapter with the same 
order in mind. We begin in section 2 by presenting a general discussion on the motion 
of planets in binary stars and their stability. Section 3, has to do with the stability 
of terrestrial planets, and in section 4, we discuss habitability and the formation of 
potentially habitable planets in a binary-planetary system\footnote{The phrase 
"binary-planetary system" is used to identify binary star systems in which one of the 
stars is host to a giant planet.}.

\begin{table}
\centering
\caption{Planets in double stars \cite{Rag06}}
\label{tab:1}
\begin{tabular}{crrcr}
\hline\noalign{\smallskip}
Star & \hskip 10pt $a_{b}$ (AU) & \hskip 10pt $a_{p}$ (AU) & 
\hskip 10pt  $M_{p} \sin i (M_{Jup})$ & \hskip 10pt $e_{p}$\\
\noalign{\smallskip}\hline\noalign{\smallskip}
HD38529 & 12042 & 0.129 & 0.78 & 0.29 \\
 & & 3.68 & 12.7 & 0.36 \\
HD40979 & 6394 & 0.811 & 3.32  & 0.23\\
HD222582 & 4746 & 1.35 & 5.11 & 0.76 \\
HD147513 & 4451 & 1.26 & 1.00 & 0.52 \\
HD213240 & 3909 & 2.03 & 4.5 & 0.45 \\
Gl 777 A & 2846 &0.128 & 0.057 & 0.1 \\ 
& & 3.92 &1.502 & 0.36\\
HD89744 & 2456 & 0.89 & 7.99 & 0.67 \\
GJ 893.2 & 2248 & 0.3 & 2.9 & -- \\
HD80606 & 1203 & 0.439 & 3.41 & 0.927 \\
55 Cnc & 1050 & 0.038 & 0.045 & 0.174 \\
 & & 0.115 & 0.784 & 0.02 \\
 & & 0.24 & 0.217 & 0.44\\
& & 5.25 & 3.92 & 0.327\\
GJ 81.1 & 1010 & 0.229 &0.11 & 0.15 \\
& & 3.167 &0.7 & 0.3 \\
16 Cyg B & 860 & 1.66 & 1.69 & 0.67\\
HD142022 & 794 & 2.8 & 4.4 & 0.57 \\
HD178911 & 789 & 0.32 & 6.292 & 0.124 \\
Ups And & 702 &0.059 & 0.69 & 0.012 \\
& & 0.83 & 1.89 &0.28 \\
& & 2.53 &3.75 & 0.27\\
HD188015 &684 & 1.19 & 1.26 & 0.15 \\
HD178911 & 640 & 0.32 & 6.29 & 0.124\\
HD75289 & 621 & 0.046 & 0.42 & 0.054 \\
GJ 429 & 515 & 0.119 & 0.122 & 0.05 \\
HD196050 & 510 & 2.5 & 3.00 & 0.28 \\
HD46375 & 314 & 0.041 & 0.249 & 0.04 \\
HD114729 & 282 & 2.08 & 0.82 & 0.31 \\
$\epsilon$ Ret & 251 & 1.18 & 1.28 & 0.07 \\ 
HD142 & 138 &0.98 &1.00 & 0.38 \\
HD114762 & 132 & 0.3 & 11.02 & 0.25 \\
HD195019 & 131 & 0.14 & 3.43 & 0.05 \\
GJ 128 & 56 & 1.30 & 2.00 & 0.2 \\
HD120136 & 45 & 0.05 & 4.13 & 0.01 \\
\hline
$\gamma$ Cep & 20.3 & 2.03 & 1.59 & 0.2\\
GL 86 & 21  & 0.11 & 4.01 &0.046  \\
HD41004 AB &  23 & 1.7 & 2.64 & 0.5 \\
\hline  
\noalign{\smallskip}\hline
\end{tabular}
\end{table}

\section{Planetary Motion in Binary Systems and Stability}

As mentioned earlier, dynamical stability is essential to the habitability of a planetary
system. This issue is particularly important in binary star systems since the gravitational perturbation of the stellar companion limits stable planetary orbits around the other 
star to only certain regions of the phase-space. In this section, we discuss this issue 
in more detail.

Study of the stability of small bodies in binary stars has a long history in planetary 
science. Among some of the early works are the papers by \cite{Gra81}, \cite{Bla82}, 
and \cite{Black83}, where the authors studied the stability of a planet around a star 
of a binary within the framework of a general three-body system, and showed that in 
a binary with equal-mass stars, the orbital stability of the planet is independent of 
its orbital inclination (also see paper by \cite{Har77}, and articles by \cite{Inn97} and
\cite{Mus05} for more recent works on this subject). 

The notion of stability has been discussed by different authors within different 
contexts. In a review article in 1984, Szebehely introduced 50 definitions for the 
stability of a planetary orbit in a multi-body system \cite{Szebehely84}. For instance, 
while \cite{Har77} considers an orbit stable if the semimajor axis and eccentricity of 
the object do not undergo secular changes, \cite{Szeb80} and \cite{Szebehely81}
define orbital stability based on the integrals of motion and curves of zero velocity. 
Within the context of habitability, an object is stable if it has the capability of 
maintaining its orbital parameters (i.e., semimajor axis, eccentricity, and inclination) 
at all times. In other words, an object is stable if small variations in its orbital
parameters do not progress exponentially, but instead vary sinusoidally. Instability 
occurs when a perturbative force causes drastic changes in the orbital parameters  
of the object so that it leaves the gravitational field of the system, or collides with 
other bodies.

\begin{figure}
\begin{center}
\includegraphics{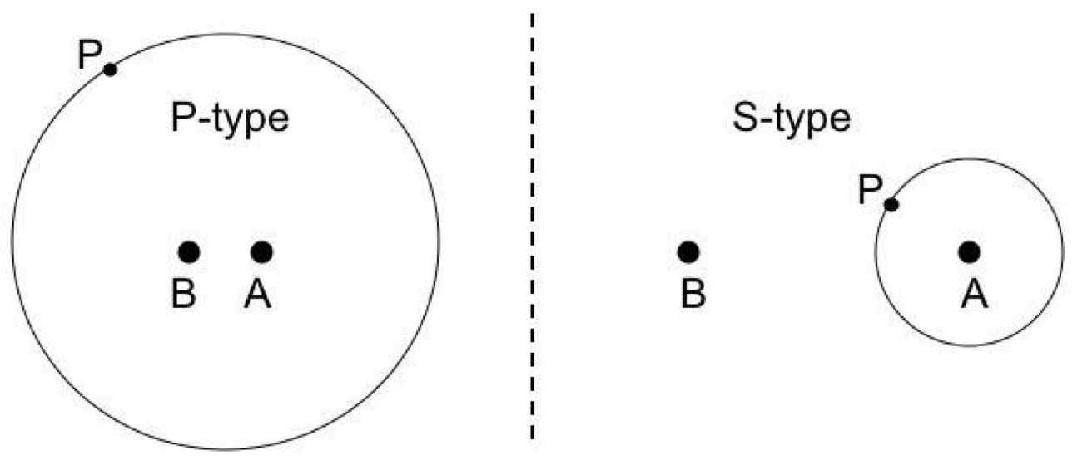}
\end{center}
\caption{S-type and P-type binary-planetary systems. A and B represent the
stars of the binary, and P depicts the planet.}
\label{fig:4}
\end{figure}

\begin{figure}
\begin{center}
\includegraphics{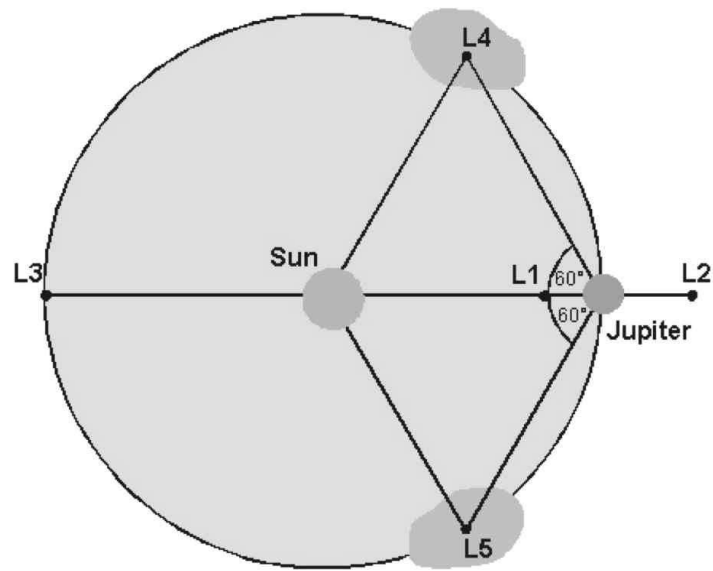}
\end{center}
\caption{Schematic view of the stable region around the Lagrange points $L_4$
  and $L_5$ in the restricted three body problem with the Sun and Jupiter as
  primary bodies}
\label{lagrange}
\end{figure}

The stability of a planetary orbit in dual-star systems depends also on the type of 
its orbit. From a dynamical point of view, three types of motion are recognized
in double-star systems (see figures \ref{fig:4} and \ref{lagrange}, and \cite{Dvo84}):

\begin{enumerate} 

\item the S-type (or the satellite-type), where the planet moves around one 
stellar component,

\item the P-type (or the planet-type), where the planet surrounds both stars in 
a distant orbit, and

\item the L-type (or the libration-type), where the planet moves in the same orbit 
as the secondary (i.e., locked in a 1:1 mean motion resonance), but $60^{\circ}$ 
ahead or behind.\footnote{An earlier classification by \cite{Szeb80} divides
the planetary orbits in binary systems into three categories: {\it inner} orbit,
where the planet orbits the primary star, {\it satellite} orbit, where it orbits the 
secondary star, and the {\it outer} orbit, where the planet orbits the entire binary 
system.}

\end{enumerate}

Within the framework of elliptical restricted three-body problem 
(ER3BP)\footnote{In an elliptical restricted three-body problem, the planet is considered 
to be a massless particle and its motion is studied in the gravitational field of 
two massive stars. The stars of the binary revolve around their center of mass in
an unperturbed elliptical Keplerian orbit.}, many authors have studied the stability 
of planets in binaries for different types of above-mentioned planetary orbits 
\cite{Dvo84,Dvo86,Rab88,Dvo89,Dvo91,Loh93,Ben88a,Ben88b,Ben89,Ben93,
Ben96,Ben98, Holman97,Hol99,Pil02}. However, because until 2003, no planet 
had been detected in or around a double star, the applicability of the results of these 
studies were only to hypothetical systems. The discovery of the first planet in a 
moderately close binary by \cite{Hat03} changed this trend
and encouraged many researcher to revisit 
this problem and explore the stability of planets in binaries by considering more 
realistic cases (see, for instance, \cite{Pil03, Dvo03a,Dvo03b,Dvo04,Hag06,Inn97,Mus05}).

In this chapter, we focus on the stability and habitability of planets in S-type orbits. 
As shown in Table \ref{tab:1}, all the currently known planets in binary systems (regardless 
of the separation of the binary) are of this kind. We present the results of the studies
of the general stability of S-type orbits and discuss their application to real binary systems, in 
particular the system of $\gamma$ Cephei. Since the discovery of a giant planet around the
primary of this double star \cite{Hat03}, many studies have been done on the stability and 
habitability of this binary and the possibility of the formation of giant and Earth-like 
planets around its stellar components \cite{Dvo03a,The04,Hag05,Ver06,Tor07}. 
We finish this section by briefly reviewing the stability of planets in P-type and L-type orbits.

The numerical simulations of planetary orbits presented in this section are mostly carried out
within the framework of the elliptical, restricted, three-body system, 
where the planet is regarded as a massless object with no influence on the
dynamics of the binary. To determine the {\it character} of the motion of 
an orbit, we either use a chaos indicator, or carry out long-term 
orbital integrations. As a chaos indicator, we use the {\it fast Lyapunov indicator} (FLI) 
as developed by \cite{Fro97}. FLI can distinguish between regular and chaotic motions 
in a short time, and chaotic orbits can be found very quickly because of the exponential 
growth of this vector in the chaotic region. For most chaotic orbits only a few 
number of primary revolutions is needed to determine the orbital behavior. 
In order to distinguish between stable and chaotic motions, we define a 
critical value for FLI which depends on the computation time. This method has 
been applied to the studies of many extrasolar planetary systems by
\cite{Pil02,Dvo03a,Dvo03b,Pil03,Boi03,Erd03,Pil05,San06}.

When carrying out long-term orbital integrations, a fast and reliable characterization of 
the motion can be achieved by making maps of the maximum eccentricity of the
orbit of the planet calculated for each integration of its orbit.
The maximum eccentricity maps can be used as a useful indicator of orbital
stability, especially for studies of the motion of terrestrial-size
planets in the habitable zone of their host stars. Examples of such studies
can be found in the works of \cite{Dvo03a,Fun04,Erd04,Dvo04,Asg04}, and \cite{Pil06}.

\subsection {Stability of S-type orbits}

The motion of a planet in an S-type orbit is governed by the
gravitational force of its host star and the perturbative effect of the binary companion.
Since the latter is a function of the distance between the planet and the secondary 
star, the orbit of the planet will be less perturbed if this distance is large. In other words, 
a planet in an S-type orbit will be able to maintain its orbit for a long
time if it is sufficiently close to its parent star \cite{Har77}. 
By numerically integrating the motion of a massless object in an S-type orbit,
\cite{Rab88} (hereafter RD) and \cite{Hol99} (hereafter HW)
have shown that the maximum value of the semimajor axis of a stable S-type 
orbit varies with the binary mass-ratio, semimajor axis, and eccentricity as,

\begin{eqnarray}
&\!\!\!\!\!\!\!\!\!\!\!\!\!\!\!\!\!\!\!\!\!\!\!\!
{{a_c}/{a_b}}=(0.464\pm 0.006)+ (-0.380 \pm 0.010)\mu 
+ (-0.631\pm0.034) {e_b} \nonumber \\
&\qquad+(0.586 \pm 0.061) \mu {e_b}
+ (0.150 \pm 0.041) {e_b^2} 
+(-0.198 \pm 0.047)\mu {e_b^2}\>.
\end{eqnarray}

\begin{figure}
\centering
\includegraphics{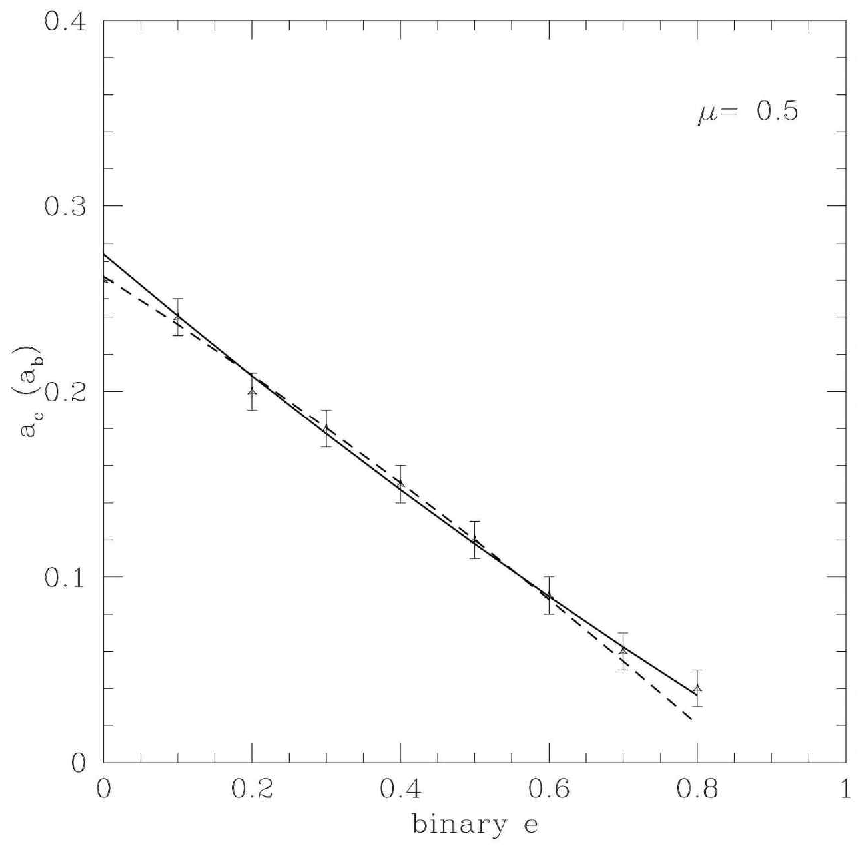}
\includegraphics{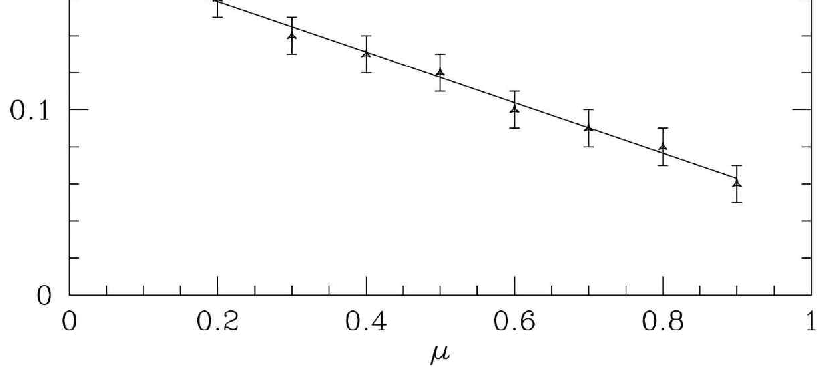}
\vskip -9.2in
\caption{Graphs of the critical semimajor axis $(a_c)$
of an S-type binary-planetary system, in units of 
the binary semimajor axis \cite{Hol99}. The graph on the top
shows $a_c$ as a function of the binary eccentricity for an equal-mass 
binary. The graph on the bottom corresponds to the variations of
the critical semimajor axis of a binary with an eccentricity of
0.5 in term of the binary's mass-ratio. The solid and dashed line on
the top panel depict the empirical formulae as reported by
\cite{Hol99} and \cite{Rab88}, respectively.}
\label{fig:5}
\end{figure}

\noindent
In this equation, $a_c$, the {\it critical} semimajor axis, is the upper limit
of the semimajor axis of a stable S-type orbit, $a_b$ and $e_b$
are the semimajor axis and eccentricity of the binary, and
$\mu={{M_2}/{({M_1}+{M_2}})}$, where 
$M_1$ and $M_2$ are the masses of the primary and secondary stars,
respectively. Figure \ref{fig:5} shows the variation of $a_c$ with the binary
mass-ratio and eccentricity. As expected, S-type orbits in binaries with
larger secondary stars on high eccentricities are less stable.
The $\pm$ signs in equation (1) define a lower and an upper value\footnote{Orbits 
with semimajor axes smaller than the lower value or larger than the upper value 
are certainly unstable}  for the critical semimajor axis which correspond to a transitional 
region that consists of a mix of stable and unstable orbits.
Such a dynamically {\it gray} area, in which the
state of a system changes from stability to instability, 
is known to exist in multi-body environments, and is a characteristic 
of any dynamical system. Similar studies have been done
by \cite{Moriwaki04} and \cite{Fatuzzo06} who obtained
critical semimajor axes slightly larger than given by
equation (1).

It is necessary to mention that in simulations of RD and HW,
the initial orbit of the planet was considered to be circular.
In a series of numerical integrations, \cite{Pil02} (hereafter PLD) 
considered non-zero values for the initial eccentricity of the planet and
by assuming the following initial conditions, they
analyzed the influence of the planet's eccentricity on its orbital stability. 

\vskip 5pt
\noindent
For the binary, these authors assumed
\begin{itemize}

\item a semimajor axis of 1 AU,
\item an eccentricity between 0 and 0.9 in steps of 0.1, and
\item an initial starting point for the secondary star at either its periastron or apastron. 

\end{itemize}

\vskip 5pt
\noindent
For the planet, which moves around  the primary in the same plane as the
orbit of the binary (i.e., coplanar orbits), they considered

\begin{itemize}

\item  a semimajor axis between 0.1 AU and 0.9 AU,
\item  an initial eccentricity between 0 and 0.5 in increments of 0.1 
for all binary mass-ratios, and
\item a starting point with different angular positions 
(i.e.\ mean anomaly $=0^\circ$ or $90^\circ$ or $180^\circ$ or $270^\circ$),

\end{itemize}

\begin{figure}
\begin{center}
\includegraphics{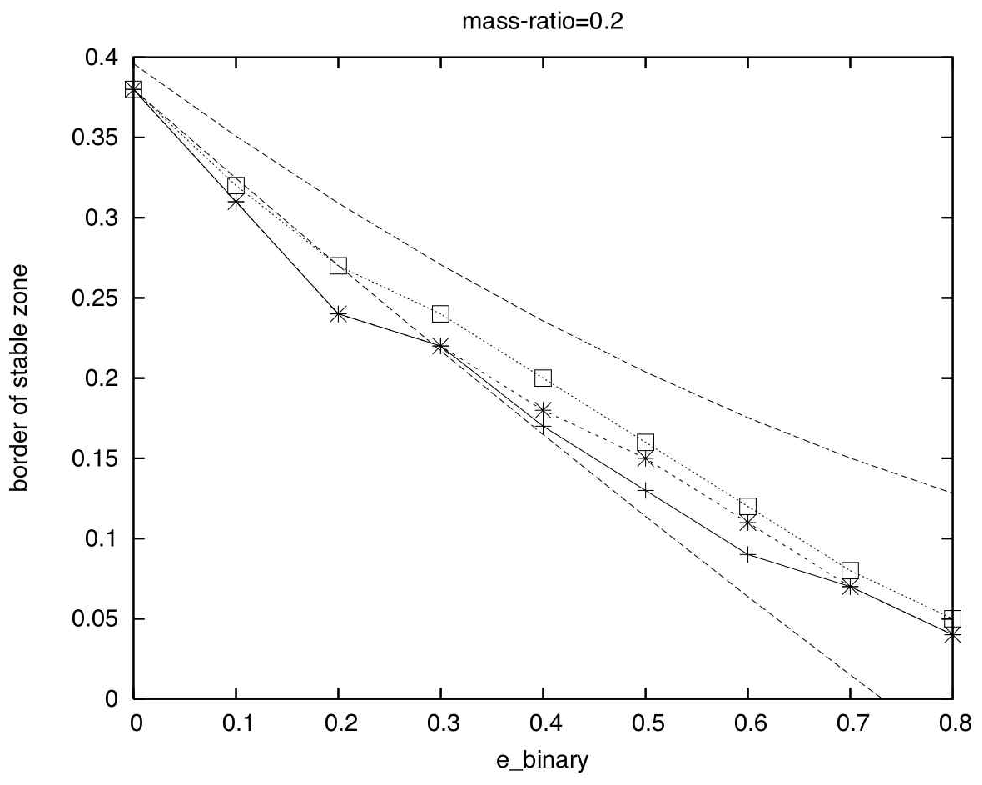}
\includegraphics{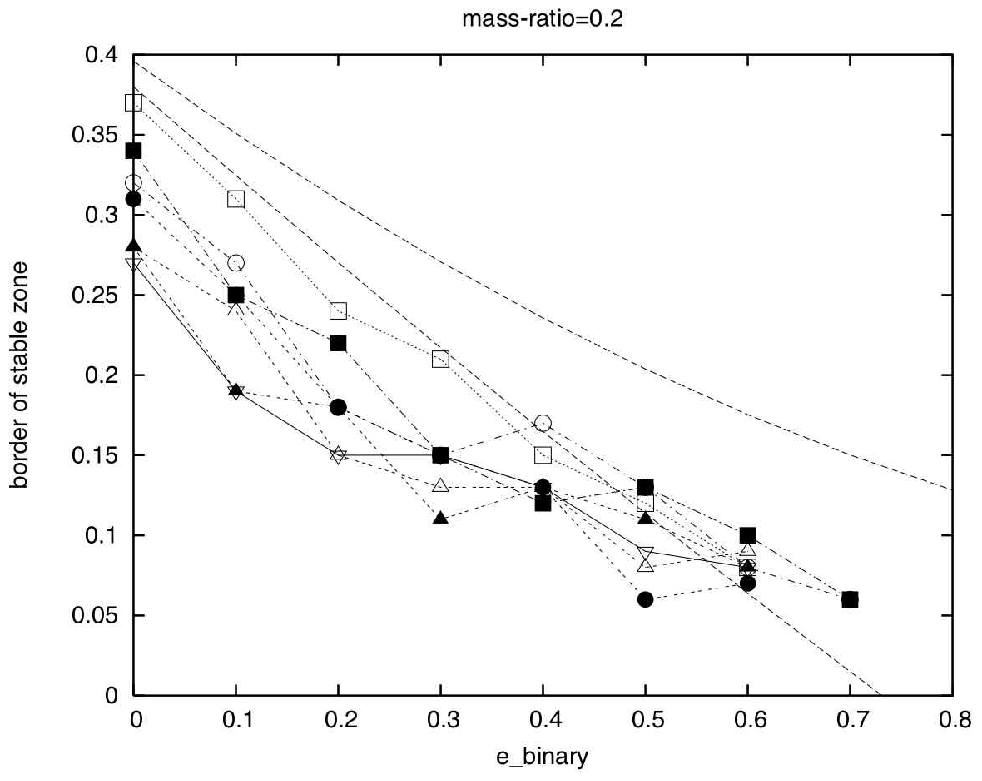}
\end{center}
\vskip -5.5in
\caption{A comparison of the results of PLD with those of HW.
The upper panel shows the results for $e_{p}=0$ (solid line with crosses) 
and $e_{p}=0.1$ (dashed line with stars) as obtained by PLD. The 
dotted line with white squares shows the results obtained by HW.
The area between the two dashed lines defines the boundary of the stable zone
according to equation (1).
The lower panel shows the results of PLD for $e_{p}$ ranging from 0.3 to 0.9.}
\label{mu02e}
\end{figure}

Figure \ref{mu02e} shows a comparison of the results of the three studies by
RD, HW, and PLD, in a binary with a mass-ratio of $\mu=0.2$. 
In this figure, the value of the critical semimajor axis 
of the planet is shown for different values of its eccentricity and the eccentricity 
of the binary. The boundaries of the stability zone 
corresponding to HW simulations (calculated using equation 1)
are shown in dotted lines. As shown in the top panel of this figure, stability zones of 
low-eccentricity orbits, as obtained by PLD, are in a good agreement with
the results of HW. However, for planets with larger orbital eccentricities,
as shown in the lower panel, the size of the stability zone decreases
as the eccentricity of the planet increase. The plotted stability boundaries 
for such orbits fall outside the HW stable zone and are closer to the planet-hosting star. 
This can also be seen in Table~\ref{tab:2}, where the lesser of the values of the inner 
boundary of the stable region (i.e.\ the semimajor axis of the last stable orbit) 
as obtained by HW and PLD, has been recorded. These results indicate that the 
stability criteria presented by HW are not applicable to eccentric S-type orbits.

It is necessary to mention that as oppose to RD and HW who determined the 
stable zone of a planet by identifying its escaping orbits within a 
certain computation time, PLD used a chaos indicator to characterize 
the long-term behavior of the planet's motion. Although because of the application of FLI, 
the computation time in PLD was much shorter than in RD and HW, 
their results are, however, valid for much longer times. In some cases in
simulations by PLD, the application of FLI resulted in a slightly larger stable 
region compare to that of HW. This is due to the fact that, as oppose to the latter,
in which 8 starting points were used, PLD used only 4 starting positions. 
Test-computations, using a different grid for the FLI-maps, and for
computation times over  $10^4, 10^5$ and $10^6$ periods
of the binary were also carried out. However, they did not change the result
significantly.

\begin{table}
\centering
\caption{Stable zone (in units of the binary semimajor axis) of an S-type orbit for 
different values of the mass-ratio and eccentricity of the binary. The given size 
for each $(\mu, {e_b})$ pair is the lesser of the values obtained by HW and PLD.}
\label{tab:2}
\hskip 10 pt  Mass-ratio $(\mu)$ \\
\begin{tabular}{cccccccccccccccccccccc}
\hline\noalign{\smallskip}
$e_b$ & & & &  0.1 & & 0.2 & & 0.3 & & 0.4 & & 0.5 & & 0.6 & & 0.7 & & 0.8 & & 0.9 \\
\noalign{\smallskip}\hline\noalign{\smallskip}
0.0 & & & & 0.45 & & 0.38 & & 0.37 & & 0.30 & & 0.26 & & 0.23 & & 0.20 & & 0.16 & & 0.13 \\ 
0.1 & & & & 0.37 & & 0.32 & & 0.29 & & 0.27 & & 0.24 & & 0.20 & & 0.18 & & 0.15 & & 0.11 \\
0.2 & & & & 0.32 & & 0.27 & & 0.25 & & 0.22 & & 0.19 & & 0.18 & & 0.16 & & 0.13 & & 0.10 \\
0.3 & & & & 0.28 & & 0.24 & & 0.21 & & 0.18 & & 0.16 & & 0.15 & & 0.13 & & 0.11 & & 0.09 \\
0.4 & & & & 0.21 & & 0.20 & & 0.18 & & 0.16 & & 0.15 & & 0.12 & & 0.11 & & 0.10 & & 0.07 \\
0.5 & & & & 0.17 & & 0.16 & & 0.13 & & 0.12 & & 0.12 & & 0.09 & & 0.09 & & 0.07 & & 0.06 \\
0.6 & & & & 0.13 & & 0.12 & & 0.11 & & 0.10 & & 0.08 & & 0.08 & & 0.07 & & 0.06 & & 0.045 \\
0.7 & & & & 0.09 & & 0.08 & & 0.07 & & 0.07 & & 0.05 & & 0.05 & & 0.05 & & 0.045 & & 0.035 \\
0.8 & & & & 0.05 & & 0.05 & & 0.04 & & 0.04 & & 0.03 & & 0.035 & & 0.03 & & 0.025 & & 0.02 \\
\noalign{\smallskip}\hline
\end{tabular}
\end{table}

Table \ref{tab:3} shows the variations of the size of the stable zone in simulations
of PLD in terms of the eccentricities of the binary and planet, and for
different binary mass-ratios. As shown here, as the eccentricity of the binary
increases, the boundary of the stable zone varies from 0.04 (for an initially eccentric 
motion in a binary with an eccentricity of 0.5 and mass-ratio of $\mu=0.9$) to 
0.45 (for an initially circular motion in a circular binary with $\mu=0.1$). 
Table \ref{tab:3} also shows that the size of the stable region does not have a strong
dependence on the eccentricity of the planet. This dependence is not,
however, negligible, especially if a planet is close to the border of the 
chaotic motion and moves in a highly eccentric orbit. 
A presentation of the 3-D stability plots for different mass-ratios with a detailed
discussion can be found in PLD.

\begin{table}
\centering
\caption{Stable zone (normalized to $a_{b}$) of an S-type orbit}
\label{tab:3}
\hskip 80 pt  Stable Zone \\
\begin{tabular}{cccccccccc}
\hline\noalign{\smallskip}
Mass-ratio $(\mu)$& $e_b$ & & & & ${e_p}=0$ & & & &${e_p}=0.5$ \\
\noalign{\smallskip}\hline\noalign{\smallskip}
0.1&  0 & &  & & 0.45 & & & & 0.36 \\
& 0.5 &   & & & 0.18 & & & & 0.13 \\
\hline
0.2 & 0 & & & & 0.40 & & & & 0.31 \\
& 0.5 & & & & 0.16 & & & & 0.12 \\
\hline
0.3 & 0 & & & & 0.37 &  & & & 0.28 \\
& 0.5 & & & & 0.14 & & & & 0.11 \\
\hline
0.4 & 0 & & & & 0.30 & & & &0.25 \\
& 0.5 & & & & 0.12 & & & & 0.07  \\
\hline
0.5 & 0 & & & & 0.27 & & & & 0.22\\
& 0.5 &  & & & 0.12 & &  & & 0.07\\
\hline
0.6 & 0 & & & & 0.23 & & & & 0.21 \\
 & 0.5 & & & & 0.10 & & & & 0.07 \\
\hline
0.7 & 0 & & & & 0.20 & & & & 0.18 \\
& 0.5 & & & & 0.09 & & & & 0.07 \\
\hline
0.8 & 0 & & & & 0.16 & & & & 0.16 \\
& 0.5 & & & & 0.09 & & & & 0.05 \\
\hline
0.9 & 0 & & & & 0.13 & & & & 0.12 \\
 & 0.5 & & & & 0.06 & & & & 0.04 \\
\noalign{\smallskip}\hline
\end{tabular}
\end{table}

An interesting application of the analysis of HW and PLD is to the stability
of terrestrial planets and smaller objects. Since in the calculations of
the critical semimajor axis by these authors, a giant planet was consider to be a test particle,
given that the mass of a Jovian-type planet is approximately two
orders of magnitude larger than the mass of a terrestrial-class object, the stability criteria
of HW and PLD can be readily generalized to identify regions around the stars
of a binary where terrestrial-class planets can have long-term stable orbits
\cite{Quintana02,Quintana06,Quintana07}. This results can also be used to
identify regions where smaller objects, such as asteroids, comets, and/or
dust particles may reside. Although these analyses do not
include non-gravitational forces, their applications to observational
data has been successful and have identified dust bands, possibly due to
the collision among planetesimals, in several wide S-type binaries
(see figures 8 and 9 of chapter 1, and  \cite{Trilling07}).

\subsubsection{Application to the binary $\gamma$ Cephei}

Gamma Cephei is one of the most interesting double star systems that host 
a planet. At a distance of approximately 11 pc from the Sun, and with a semimajor
axis and an eccentricity of 18.5 AU and 0.36, respectively, this system
present a prime example of a moderately close binary with a planet in an
S-type orbit. The primary of $\gamma$ Cephei, a 1.6 solar-mass K1 IV sub-giant
\cite{Fuhr04} is host to a stellar companion, an M4 V star with a mass of 
0.4 solar-masses \cite{Neu07,Tor07}, and a Jovian-type planet with
a mass of 1.7 Jupiter-mass and an eccentricity of 0.12 at 1.95 AU \cite{Hat03}. 

\begin{figure}
\vskip 1in
\begin{center}
\includegraphics{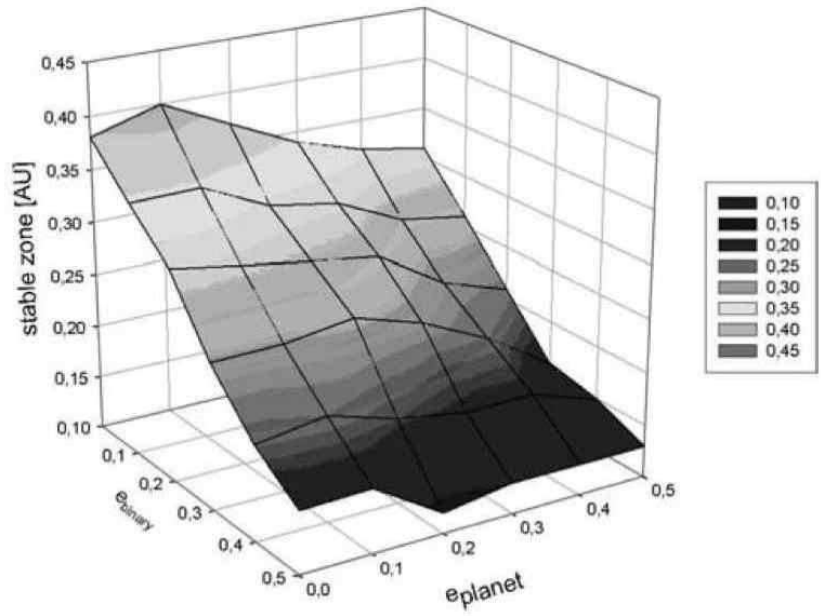}
\end{center}
\caption{The stable zone of an S-type orbit in a binary with 
mass-ratio $\mu=0.2$ (e.g., $\gamma$ Cephei).
As shown here, the extent of the stable zone is more
strongly affected by the eccentricity of the binary than that of the planet \cite{Pilat04}.}
\label{mu02}
\end{figure}

\begin{figure}
\begin{center}
\vskip 1.7in
\includegraphics{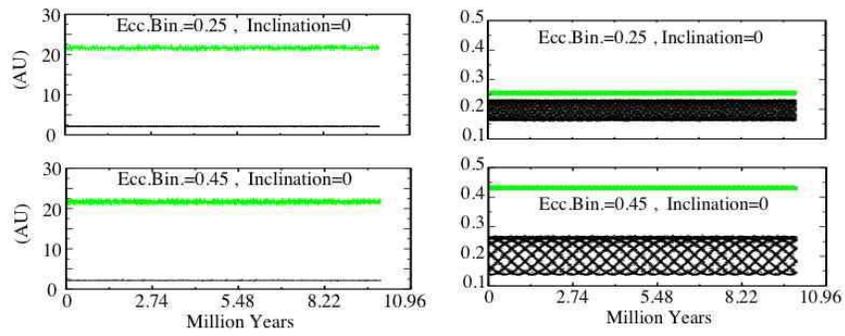}
\end{center}
\caption{Graphs of the semimajor axes (left) and eccentricities (right) 
of the giant planet (black) and binary (green) of $\gamma$ Cephei
for different values of the eccentricities of the binary \cite{Hagh04}.
The mass-ratio of the binary is 0.2.}
\label{fig:6}
\end{figure}

\begin{figure}
\vskip 2in
\begin{center}
\includegraphics{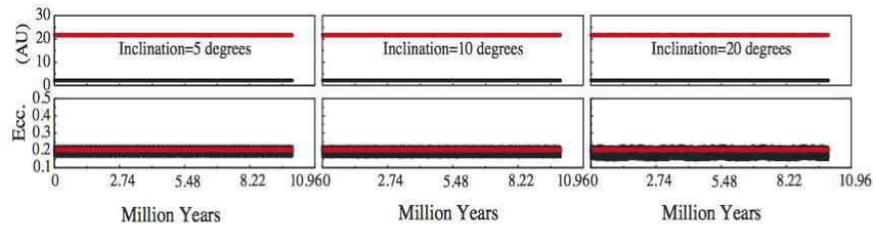}
\end{center}
\caption{Graphs of the semimajor axes (top) and eccentricities (bottom) 
of the giant planet (black) and binary (red) of $\gamma$ Cephei. 
The initial eccentricity of the binary at the beginning of numerical 
integration and the value of its mass-ratio were equal to 0.20 \cite{Hag06}.}
\label{fig:7}
\end{figure}

The mass-ratio of $\gamma$ Cephei binary is 0.2 making this system a suitable example 
for applying the stability analysis of S-type orbits as discussed in the previous section. 
An overview of the size of the stable region for the giant planet of this system
is shown in figure \ref{mu02} where the planet maintained its orbit for 1000 time units.  
As shown in this figure, the zone of stability for the giant planet extends to approximately 
3.16 AU\footnote{Test-computations for $\mu=0.3, 0.5$ and 0.7, up to 100,000 time 
units showed the same qualitative results.}.
Direct integration of the binary and the planet for different values of the binary eccentricity 
indicate that the orbit of the planet is stable when $0.2 \leq {e_b} \leq 0.45$ \cite{Hag06}. 
Samples of the results of these integrations are shown in figure \ref{fig:6}.
Numerical integrations of the orbit of the planet for different values of its inclination 
with respect to the plane of the binary $(i_p)$ show that this object is stable for 
inclinations less than 40$^\circ$. Figure \ref{fig:7} shows the semimajor axes and 
orbital eccentricities of the system for ${e_b}=0.2$ and for  $i_p$=5$^\circ$, 10$^\circ$,
and 20$^\circ$.

\begin{figure}
\vskip 1.7in
\begin{center}
\includegraphics{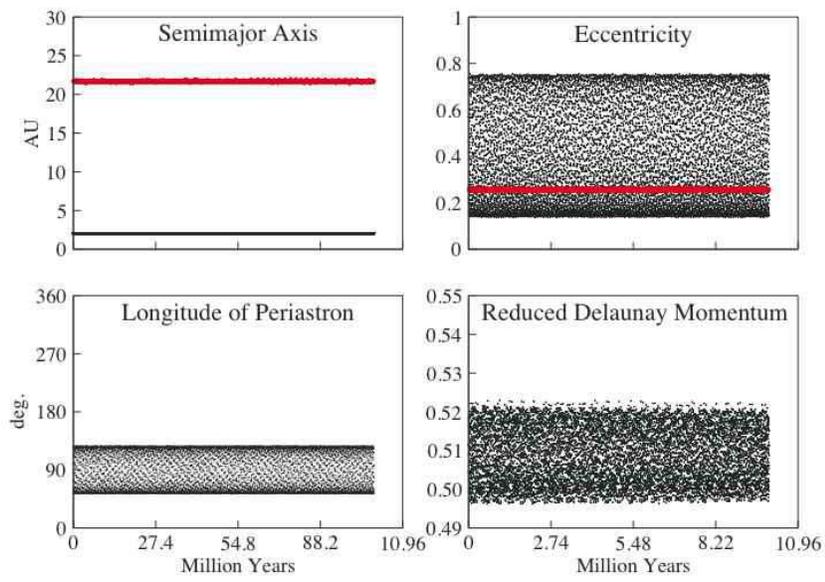}
\end{center}
\caption{Graphs of the semimajor axis and eccentricity of the giant planet
(black) and binary (red) of $\gamma$ Cephei (top) and its
longitude of periastron and reduced Delaunay momentum (bottom)
in a Kozai resonance \cite{Hagh04,Hagh05a}
. As expected, the longitude of the periastron of the
giant planet oscillates around $90^\circ$ and its reduced Delaunay
momentum is constant.}
\label{fig:8}
\end{figure}

\begin{figure}
\begin{center}
\includegraphics{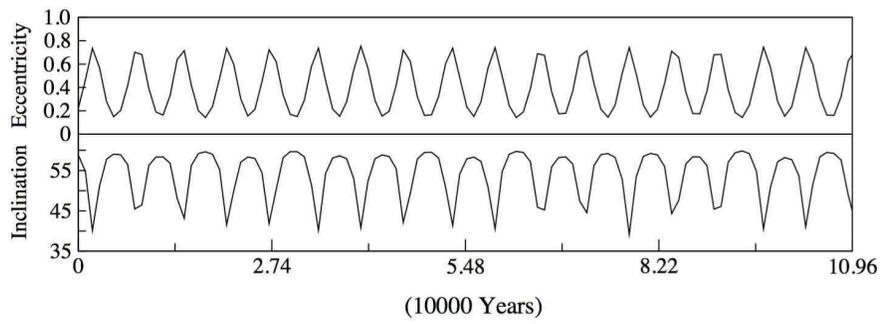}
\end{center}
\caption{Graphs of the eccentricity and inclination of the giant
planet of $\gamma$ Cephei in a Kozai Resonance \cite{Hagh04,Hagh05a}. As
expected, these quantities have similar periodicity and are 
$180^\circ$ out of phase.}
\label{fig:9}
\end{figure}

Numerical simulations also indicated the possibility of a Kozai resonance in
the $\gamma$ Cephei system. Kozai resonance has been studied in binary-planetary
systems by several authors \cite{Hagh04,Hagh05a,Ver06,Takeda06,Malmberg07,Takeda08,
Saleh09}.  As demonstrated by \cite{Kozai62}, in a three-body 
system with two massive bodies and a small object, such as an S-type binary-planetary system,
the orbital eccentricity of the planet can reach high values at large inclinations
due to the exchange of angular momentum between the planet and the secondary star.
In such cases, the longitude of the periastron of the planet, $\omega_p$, librates
around a fix value. Figure \ref{fig:8} shows this for the giant
planet of $\gamma$ Cephei. As shown here, $\omega_p$ librates 
around 90$^\circ$ \cite{Hagh04,Hagh05a}. The inclination of the planet of $\gamma$ Cephei, 
when in a Kozai resonance, is related to its longitude of periastron and  
orbital eccentricity $(e_p)$ as \cite{Inn97}
\begin{equation}
{\sin^2}{\omega_p}=\,0.4\,{\csc^2}{i_p},
\end{equation}
\noindent
and

\begin{equation}
{(e_p^2)_{\rm max}}={1\over 6}\,\Bigl[1-5\cos (2{i_p})\Bigr].
\end{equation}
\vskip 10pt
\noindent
Equation (2) indicates that the Kozai resonance may occur
if the orbital inclination of the small body is
larger than 39.23$^\circ$. For instance, as shown by \cite{Hagh04,Hagh05a}, 
in the system of $\gamma$ Cephei, Kozai resonance occurs 
at ${i_p}={60^\circ}$.
For the minimum value of ${i_p}$, the maximum value
of the planet's orbital eccentricity is reached and, as given by equation (3), 
is equal to 0.764. Figures \ref{fig:8} and \ref{fig:9} show that
$e_p$ stays below this limiting value at all times.

\subsubsection{Application to binaries Gliese 86 and HD41004}

The application of the stability analysis of section 2.1 to the planet of
the binary Gliese 86 indicates that the orbit of this planet is stable.
This is not surprising since with a semimajor axis of 0.11 AU,
this planet is close enough to the primary star to be immune from
the perturbation of the other stellar companion.

In the case of HD41004, the application of the stability analysis of section 2.1 
is not straightforward; the orbital parameters of the planet in this binary has not been uniquely
determined. The value of the semimajor axis of this planet varies between 1.31 AU 
and 1.7 AU, and its orbital eccentricity seems to be quite high (between 0.39 and 0.74)
\cite{Zucker04}. Since the eccentricity of the binary HD41004 is unknown, 
to determine the stable zone of this system, simulations were carried out  for 
different sets of orbital parameters as a function of the binary eccentricity. 
The results indicate that the stability of 
the planet is strongly correlated with its orbital eccentricity and the eccentricity 
of the binary. Simulations show that in all cases, in order to obtain stability,
binary eccentricity has to be smaller than 0.6. For high values of the planet's
eccentricity (e.g., 0.74), the value of the binary eccentricity has to become
even smaller (less than 0.15) to ensure that the orbit of the planet will stay stable.

\subsection{Stability of P-type Orbits}

Although, no circumbinary planet has yet been discovered,
stability of P-type orbits has been a subject of research for many years
\cite{Ziglin75,Szebehely81,Dvo84,Dvo86,Dvo89,Kubala93,
Hol99,Broucke01,Pil03,Mus05}. In general, a planet in a
P-type orbit is stable if its distance from the
binary is so large that the perturbations of the binary stars
cannot disturb its motion. Such a stable planet cannot, however, 
orbit the binary too far from its center of mass since galactic perturbations
and the effects of passing stars may make the orbit of the planet unstable. 
As shown by \cite{Dvo84}, for circular binaries, this distance
is approximately twice the separation of the binary, and for eccentric
binaries (with eccentricities up to 0.7) the stable region extends to
four time the binary separation.
Subsequent studies by \cite{Dvo86,Dvo89} and \cite{Hol99} have shown 
that Dvorak's 1984 results can be formulated by
introducing a critical semimajor axis
below which the orbit of the planet will be unstable;

\begin{eqnarray}
&\!\!\!\!\!\!\!\!\!\!\!\!\!\!\!\!\!\!\!\!\!\!\!\!\!\!
{a_c}/{a_b}=(1.60 \pm 0.04) + (5.10 \pm 0.05){e_b}
+ (4.12 \pm 0.09)\mu \nonumber \\
&\qquad\qquad
+ (-2.22 \pm 0.11){e_b^2} + (-4.27 \pm 0.17){e_b}\mu
+ (-5.09 \pm 0.11){\mu^2} \nonumber \\
&\!\!\!\!\!\!\!\!\!\!\!\!\!\!\!\!\!\!\!\!\!\!
\!\!\!\!\!\!\!\!\!\!\!\!\!\!\!\!\!\!\!\!\!\!\!\!\!\!\!\!\!
\!\!\!\!\!\!\!\!\!\!\!\!\!\!\!\!\!\!\!\!\!\!\!\!
+ (4.61 \pm 0.36){e_b^2}{\mu^2}\>.
\end{eqnarray}

\begin{figure}
\vskip 1.5in
\begin{center}
\includegraphics{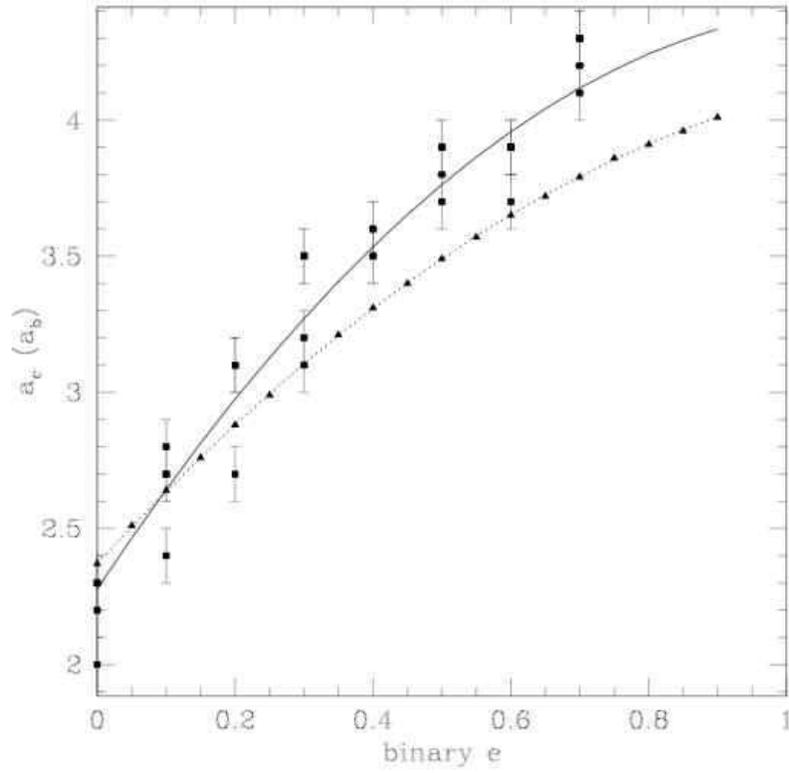}
\end{center}
\caption{Critical semimajor axis as a function of the binary eccentricity in 
a P-type system. The squares correspond to the result of 
stability simulations by \cite{Hol99} and the triangles represent those
of \cite{Dvo89}. The solid line corresponds to equation (4). As indicated
by \cite{Hol99}, the figure shows that at outer regions, the stability 
of the system fades away.}
\label{fig:10}
\end{figure}

\noindent
Figure \ref{fig:10} shows the value of $a_c$ for different values of the binary eccentricity.
Similar to S-type orbits, the $\pm$ signs in equation (4) 
define a lower and an upper value for the critical semimajor axis $a_c$, and set
a transitional region that consists of a mix of stable and unstable orbits. 
We refer the reader to \cite{Hol99,Pil03} and \cite{Pil06a} for
more details. Using this stability criteria in analysis of their observational
data, \cite{Trilling07} have been able to detect circumbinary dust bands, possibly 
resulted from the collision of planetesimal, around several close binary stars
(Chapter 1, figures 8 and 9).

A dynamically interesting feature of a circumbinary stable region 
is the appearance of islands of instability. As shown by \cite{Hol99},
islands of instability may develop beyond the inner boundary of the
{\it mixed zone}, which correspond to the locations of $(n:1)$
mean-motion resonances. The appearance of these unstable regions
have been reported by several authors
under various circumstances \cite{Henon70,Dvo84,Rab88,Dvo89}.
Extensive numerical simulations would be necessary to determine
how the overlapping of these resonances would affect the stability
of P-type binary-planetary orbits.

\subsection{Stability of L-Type Orbits}

The L-type orbit, in which an object librates around one of the binary's 
Lagrangian triangular points (figure \ref{lagrange}), may not be entirely relevant to 
planetary motions in double star systems. The reason is that such an orbital 
configuration requires ${M_2}/{({M_1}+{M_2})} \leq 1/26$, which is better
fulfilled in systems consisting of a star and a giant planet. Recent simulation
by \cite{Hagh08} have shown that in systems with a close-in giant planet, 
L-type planetary orbits with
low eccentricities can be stable for long times. We refer the reader to
section 3.2 and the paper by \cite{Pil03} for more details on the stability of these orbits.

\section{Terrestrial Planets in Binaries}

In the previous section, a general analysis of the dynamics of a planet
in a binary star system was presented. However, within the context of habitability, 
the interest falls on the motion and long-term stability of Earth-like planets. 
It would be interesting to extend studies of the habitability, similar to  those
by \cite{Jon01} and \cite{Men03}, to binary star system, in particular those in which 
a giant planet already exists, and analyze the dynamics of fictitious Earth-like planets 
in such complex environments. In this section, we focus on this issue.

In general, four different types of orbits are possible for a terrestrial planet
in a binary system that hosts a giant planet:

\begin{itemize}

\item {\bf TP-i} : the terrestrial planet is inside the orbit of the giant planet,
\item {\bf TP-o} : the terrestrial planet is outside the orbit of the giant planet, 
\item {\bf TP-t} : the terrestrial planet is a Trojan of the primary (or secondary) 
or the giant planet, 
\item {\bf TP-s} : the terrestrial planet is a satellite of the giant planet.

\end{itemize}

\noindent
In principle, the study of the stability of these orbits requires the analysis of the dynamics
of a complicated N-body system consisting of two stars,
a giant planet, and a terrestrial-class object. Except for a few special cases, 
the complexities of these systems do not allow for an analytical treatment of their dynamics,
and require extensive numerical integrations. Those special cases are:

\begin{itemize}

\item binaries with semimajor axes larger then 100 AU in which the 
secondary star is so far away from the primary (the planet-hosting star)
that its perturbative effect can be neglected \cite{Jim02}, 

\item binaries in which the giant planet has an orbit with a very small eccentricity 
(almost circular), 

\item binaries in which, compared to the masses of the other bodies, the mass of the
terrestrial planet is negligible. In these systems, within the framework of ER3BP,
one can define curves of zero-velocity, the barriers of the motion of the
fictitious terrestrial planet, using the Jacobi constant \cite{Dvo03}. 

\end{itemize}

When numerically studying the dynamics of a terrestrial planet in a binary-planetary
system, integrations have to be carried out for a vast parameter-space.
These parameters include the semimajor axes, eccentricities, and inclinations of the  
binary and the two planets, the mass-ratio of the binary,
and the ratio of the mass of the giant planet to that of its host star. 
The angular variables of the orbits of the two planets also add to these parameters. 
Although such a large parameter-space makes the numerical analysis of the dynamics
of the system complicated, numerical integrations are routinely carried out
to study the dynamics of terrestrial planets in binary-planetary systems.
The reason is that such numerical computations allows for the
investigation of the stability of many terrestrial planets and for a grid of their 
initial conditions in one or only a few simulations. Given the small size of 
these objects compared to that of a giant planet (e.g., 1/300 in case of Earth and Jupiter), 
to the zeroth order of approximations, the effect of a terrestrial planet on the motion 
of a giant planet can be ignored, and the terrestrial planet can be considered
as a test particle. This simplification makes it possible to study the stability of
thousands of possible orbits of a terrestrial-class body in one integration. 
Several studies of the dynamical evolution of a terrestrial planet
in a binary system have used this simplification and have shown that the final results are
quite similar to the results of the numerical integrations of an actual four-body
system \cite{Dvo04,Erd05}.

In the rest of this section, we present the results of the studies of the dynamics
of a terrestrial planet, focusing primarily on  TP-i and  TP-o orbits\footnote{The 
stability of TP-t and TP-s orbits has recently been studies in a few articles
by \cite{Sch07a,Sch07b,Nau02} and \cite{Dom06}.}. Since    
no extrasolar terrestrial planet has yet been discovered,
the only possible approach for a detailed dynamical analysis 
of the orbit of such an object in a binary system is to consider a 
specific extrasolar planetary system in a double star, and study the
dynamics of a fictitious terrestrial planet for different values of its orbital
elements, and those of the binary and its giant planets.
As an example, we will consider the binary-planetary system of $\gamma$ Cephei.  
For the orbital elements of this system,
we use the values given by \cite{Hag06} and \cite{Neu07}.

\subsection{Stability of {\bf TP-i} and {\bf TP-o} Orbits}

A study of the dynamics of a  {\it full} four-body system consisting of a terrestrial planet 
in a TP-i or TP-o orbit in $\gamma$ Cephei 
indicates that the orbit of this planet can only be stable in close neighborhood
of the primary star and outside the influence zone\footnote{The influence zone
of a planetary object with a mass $m_p$ around a star
with a mass $M$ is defined as the region between
$3{R_H}-{a_p}(1-{e_p})$ and $3{R_H}+{a_p}(1+{e_p})$,
where ${a_p}$ and $e_p$ are the semimajor axis and eccentricity
of the planet, and ${R_H}={a_p}(1-{e_p}){({m_p}/{3M})^{1/3}}$ is its Hill radius.}
of the giant planet (figure \ref{fig:24}, \cite{Hag06}). Integrations also show that, while
the habitable zone of $\gamma$ Cephei is unstable \cite{Hag06,Dvo03a},
it is possible for an Earth-like planet to have a stable TP-i orbit in a region between
0.3 AU and 0.8 AU from the primary star, and when its orbit is coplanar
with that of the giant planet with an inclination less than $10^\circ$.
In the region outside the orbit of the giant planet, i.e. when the terrestrial planet is in a TP-o orbit,
the perturbations from the giant planet and the secondary star affect the stability
of this object. For instance, for the values of the binary eccentricity equal to
${e_b}$=0.25, 0.35, and 0.45, the periastron of the secondary star will be 
as close as 13.9 AU, 12.0 AU, and 10.2 AU, respectively. At these distances, the 
secondary will have strong effects on the stability of the orbit of a terrestrial planet
with a semimajor axis between 2.5 AU and 5.8 AU. Simulations show that
no orbit survives in this region longer than approximately
$10^5$ years. For TP-o orbits inside 2.5 AU, the perturbation of the giant planet
is the main factor in the instability of the orbit of the terrestrial planet.

To study the effect of orbital inclination on the stability of a terrestrial planet
in $\gamma$ Cephei, the region between the host-star and the giant planet
of this system was examined for different values of the inclination of a fictitious 
terrestrial-size object, with and without the secondary star \cite{Pilat04}.
Figure \ref{gamma} shows the results. While dynamical models using two massive bodies
(i.e., primary star and the giant planet) show a vast region of stability for a massless terrestrial
planet (gray area in the lower panel of figure \ref{gamma}), 
models with three massive bodies (i.e., primary, secondary, and the giant planet),
show a decrease in the stable region (see upper panel of figure \ref{gamma}). They also show
an arc-like chaotic path with an island of stability around 1 AU, which corresponds
to the 3:1 mean motion resonance with the giant planet. 

The instability of the orbit of a terrestrial planet in $\gamma$ Cephei system (in particular
in its habitable zone) has been studied only for prograde orbits. Recently in an article 
by \cite{Gay08}, the authors investigated the stability of 
retrograde orbits in extrasolar planetary 
systems with multiple planets and showed that in systems were prograde orbits are
unstable (e.g., HD 73256), retrograde orbits may survive for long times. The stability
of retrograde orbits in planetary systems has been known
for many years \cite{Har72,Har75,Har77,Don94}. Such long-term stable orbits have also been
observed among Jupiter's retrograde irregular satellites \cite{Jewitt07}.  
To investigate whether retrograde TP-i and TP-o orbits can survive in a binary-planetary
system, the motion of a massless terrestrial planet in these orbits was simulated 
for 1 Myr in the binary of $\gamma$ Cephei. Figure \ref{TPio} shows the results. 
From this figure one can see that for semimajor axes smaller than 1.8 AU, a 
TP-i orbit is stable. However, at close distances to the giant planet, this orbit 
suffers from strong perturbations from this object and becomes unstable (straight long lines 
around 2 AU where the eccentricity of terrestrial planet reaches unity). Figure \ref{TPio}
also shows that for a TP-o orbit, a stability region exists for initial semimajor axes 
ranging from 2.5 AU to 7.2 AU, with the minimum perturbation received at the
semimajor axis of 3.2 AU. Beyond 7.2 AU, terrestrial 
planet becomes unstable due to the perturbation from the secondary star. 
The three small peaks in the stable region of figure \ref{TPio} correspond to mean-motion
resonances between the terrestrial planet and the giant planet. A comparison between this
figure and figure \ref{fig:24}, in which the lifetime of a terrestrial planet in a prograde 
circular orbit is shown, clearly indicates that $\gamma$ Cephei has a large stable region for
retrograde orbits, in particular in the habitable zone of its primary star.

Figure \ref{TPi} shows the results of the integrations of a terrestrial planet in a retrograde
TP-i orbit in the $\gamma$ Cephei system, in terms of the initial orbital inclination of this 
object. The initial semimajor axis of the terrestrial planet was varied between 
0.4 AU and 2.0 AU, and its initial orbital inclination was chosen to be between 
$135^{\circ}$ and $180^{\circ}$. The eccentricity of the binary was $e_{b}=0.35$. 
As shown in this figure, the upper stability limit for a  terrestrial planet in retrograde orbits is 
1.8 AU corresponding the inclinations between $165^{\circ}$ and $180^{\circ}$,
and the lower limit is 0.8 AU for an inclination of $145^{\circ}$. 
For more inclined orbits of a fictitious terrestrial planet, this lower limit drops to 
0.4 AU. Simulations for different values of the binary eccentricity indicate
that the direct effect of the binary orbit on the stability of a
terrestrial planet in a retrograde TP-i orbit is negligible, and it only affects the
eccentricity of the orbit of the giant planet (${\delta e}_{GP} = 0.08, 0.1, 0.11$ for 
$e_{b} = 0.25, 0.35, 0.45$, respectively). It is the latter that affects 
the orbit of the fictitious planet in a TP-i orbit.

The situation is different for a retrograde TP-o orbit.
As shown in figure \ref{TPo}, although the effect of the 2:1 resonance with the giant 
planet at 3.1-3.6 AU makes the orbit of a retrograde terrestrial planet unstable, 
large stable regions, especially for lower values of the binary eccentricity, 
exist beyond this region and for different values of the 
initial inclination of the orbit of the terrestrial planet. For instance, for $e_{b}=0.25$, 
the region of stability extends from 3.5 AU to 5.8 AU for the values of the
inclinations ranging from $145^\circ$ to $180^\circ$. At the 2:1 mean-motion resonance also 
a stable region exists for inclinations between $160^{\circ}$ and $180^{\circ}$,
when the eccentricity of the terrestrial planet is smaller than 0.2. In the case of a binary with 
$e_{b}=0.35$ (figure \ref{TPo}, middle graph), the unstable region corresponding
to the 2:1 resonance extends to higher inclinations. For the values
of the eccentricity of the binary larger than 0.45, instability extends
to almost all inclinations.

\subsection{Stability of {\bf TP-t} Orbits}

An interesting case of a stable orbit is when the terrestrial planet is in a Lagrange 
equilibrium point either at an angular separation of $60^{\circ}$ ahead of 
a giant planet or behind it (figure \ref{lagrange}).  
In the simplified dynamical model of the restricted (circular and elliptic) 
three-body system, many investigations exist concerning the stability of
such a planet in terms of the mass-ratio of the planet-hosting star and
its giant planet \cite{Rab61}, and the eccentricity of the giant planet's
orbit \cite{Dep70}. Within the context of extrasolar planetary systems,
stability of the orbit of a terrestrial planet in a Lagrangian point has been studied
by \cite{Lau02,Men03,San03,Erd04,Sch05b,Sch07a,Sch07b,Sch09}. 
Recent simulations by \cite{Hagh08} and \cite{Capen09}
show that terrestrial planets as Trojans of giant planets can also exist in systems
where the giant planet transits its host star. Figure \ref{megno} shows an example
of such systems. In this figure, a solar-type star is host to a Jupiter-size
transiting planet in a 3-day circular orbit. The graph shows the stability of an
Earth-size object for different values of its semimajor axis and
orbital eccentricity. As shown here, a terrestrial planet in a 1:1 resonance with the giant
planet can maintain a stable orbit for eccentricities ranging from 0.2
to 0.5.

To study the stability of TP-t orbits in the $\gamma$ Cephei system, 
the orbit of a terrestrial planet was integrated in a Lagrangian point of the 
giant planet and in a general four-body system. Because as shown by \cite{Sch07a}, 
in order for a small body to have a stable Trojan orbit in an elliptical, restricted,
three-body system, the orbital eccentricity of the giant planet cannot
exceed 0.3, the eccentricity of the giant planet of $\gamma$ Cephei 
was set to this value.
Figure \ref{l4-l5} shows the results. As shown here, a region of stability
exists for a Trojan terrestrial planet around the giant planet. Small variations in the
orbital eccentricity of the giant planet, which are due to the perturbation of the secondary
companion, cause the apparent asymmetry in the location of the stable orbits
around the two Lagrangian points. These regions contain very
stable orbits with eccentricities smaller than 0.2.  
The extension of semimajor axis associated
with this asymmetry is small (only 2.5\%). A transition from
an eccentricity of 0.35 for the orbits on the edge of the stable zone,
to 0.5 for orbits in the unstable region are also shown.

\section{Habitable Planet Formation in Binaries}

As seen in previous chapters, planet formation in 
close and moderately close binary star systems is an active topic of research. 
Whether models of giant and terrestrial planet formation around single stars
can be extended to binary systems depends strongly on the orbital elements 
of the binary, its mass-ratio, and the types of its stars. While the
detection of systems such as L1551 (figure \ref{fig:14})
implies that planet formation in binaries may proceed in the same fashion as around
single stars, simulations such as those by \cite{Hep78,Artymowicz94,Whitmire98} 
and \cite{Pichardo05} indicate that a stellar component in an eccentric orbit can
considerably affect planet formation by 

\begin{itemize}

\item increasing the relative velocities of  planetesimals, which may cause 
their collisions to result in breakage and fragmentation, 

\item truncating the circumprimary disk of embryos to smaller radii, 
which causes the removal of material that may be used in the formation of 
terrestrial planets (figure \ref{fig:13}), and 

\item destabilizing the regions where the building blocks of these objects may exist.

\end{itemize}

Prior to the detection of planets in binary stars, studies of planet
formation in these systems were limited to only some specific 
or hypothetical cases. For instance,  
\cite{Hep74,Hep78,Drob78,Diakov80,Whitmire98,Kortenkamp01}
studied planet formation in binaries where the system consisted of Sun and 
Jupiter, and the focus was on the effect of Jupiter on the formation of inner 
planets of our solar system. \cite{Whitmire98} studied planet formation
in binaries, in particular those resembling some of extrasolar planets, in which the 
secondary star has a mass in the brown dwarf regime. 
\cite{Barbieri02,Quintana02,Lissauer04} also studied the late stage of terrestrial
planet formation (i.e., growth of planetary embryos to terrestrial-size objects) 
in the $\alpha$ Centauri system. 

The detection of the giant planet of $\gamma$ Cephei changed this trend. 
By providing a real example of a planetary system in a binary star, this
discovery made theorist take a deeper look at the models of planet formation
and focus their efforts on explaining how this planet was formed and whether
such systems could harbor smaller planetary objects. The results of their works, however,
have made the matter quite complicated. For instance, while simulations
as those presented in chapter 10 by Quintana \& Lissauer imply that the late
stage of terrestrial planet formation may proceed successfully in binary star systems
and result in the formation of terrestrial-class objects, simulations of earlier
stages have not been able to model the accretion of planetesimals to planetary embryos. 
On the other hand, as indicated by Marzari and co-authors in chapter 7, 
despite the destructive role of the binary companion, i.e., increasing
the relative velocities of planetesimals, which causes their collisions
to result in erosion, growth of these objects to larger sizes may still
be efficient as the effect of the binary companion can be counterbalanced
by dissipative forces such as gas-drag and dynamical friction.
As shown by these authors, for planetesimals of comparable sizes,
the combined effect of gas-drag and the gravitational
force of the secondary star may result in the alignment of the periastra
of small objects and increase the efficiency of their accretion
by reducing their relative velocities \cite{Marzari97,Marzari00}. 
However, the efficiency of this mechanism depends on the size of the 
planetesimals\footnote{For colliding
bodies with different sizes, depending on the size distribution of
small objects, and the radius of each individual planetesimal,
the process of the alignment of periastra may instead increase
the relative velocities of the two objects, and cause their
collisions to become eroding \cite{Thebault06}.}, the eccentricity of the
planetesimals disk, and the orbital elements of the binary system. 
As shown by \cite{Paardekooper08}, depending on the perturbation of the secondary
star, the eccentricity of the disk may reach a limiting value below which the encounter velocities
of planetesimals are within a factor 2 of their corresponding values in a circular
disk, and above that, the encounter velocities become so high that planetesimal
accretion is inhibited. The application of these simulations to the $\alpha$ Centauri system
has shown that the growth of planetesimals to planetary embryos may be impossible within
0.5 AU  of the primary star of this system \cite{Thebault08}. However, simulations
by \cite{Thebault09} and \cite{Marzari09} indicate that this process can be efficient around
$\alpha$ Centauri B, and planetary embryos can form within the terrestrial/habitable
region of this star. Similar results have been obtained by \cite{Xie08,Xie09} when
numerically integrating a slightly inclined disk of planetesimals around the primary of
$\gamma$ Cephei. As shown by these authors, the gas-drag causes the sorting of 
inclined planetesimals according to their sizes, and increases the efficiency
of their accretion by decreasing their relative velocities. For larger values of 
planetesimals inclinations, accretion is more efficient in wide (e.g., $> 70$ AU) 
binaries \cite{Marzari09}.

As one can notice, a common starting point in all these simulations is 
after km-sized object or larger bodies have already formed.
The reason is that among the four stages of planet formation, that is,

\begin{itemize}

\item coagulation of dust particles and their growth to centimeter-sized
objects,

\item growth of centimeter-sized particles to 
kilometer-sized bodies (planetesimals),

\item formation of Moon- to Mars-sized protoplanets (also known as 
planetary embryos) through the collision and accretion of 
planetesimals, and

\item collisional growth of planetary embryos to 
terrestrial-sized objects,

\end{itemize}

\noindent
the last two can be more readily studied in a system of double stars. At
these stages, the dominant force in driving the dynamics of objects is 
their mutual interactions through their gravitational forces, 
and the simulations can be done using
N-body integrations. In chapter 10 of this volume, Quintana \& Lissauer 
have presented the results of a series of such simulations, and
investigated terrestrial planet formation in binary systems such
as $\alpha$ Centauri. Given that
the late stage of terrestrial planet formation is a slow process, 
which may take a few hundred million years, it is possible that during
the first few million years of this process, giant planets are also formed
at large distances from the planet-hosting star. Similar to terrestrial
planet formation in our solar system, these objects will play a vital role
in the formation, distribution, and water-content of terrestrial-class
objects in binary systems. Within the context of habitable planet formation,
this implies that the formation of terrestrial planets has to be
simulated while the effects of the secondary and the giant planet(s) of
the system are also taken into account.

\subsection {Habitable Zone}

Life, as we know it, requires liquid water. A potentially habitable planet
has to be able to maintain liquid water on its surface and in 
its atmosphere. The capability of a planet in maintaining water
depends on many factors such as its size, interior dynamics, atmospheric circulation,
and orbital parameters (semimajor axis and orbital eccentricity). It also depends
on the brightness of the central star at the location of the planet. 
These properties, although at the surface unrelated, have strong intrinsic correlations, 
and combined with the luminosity of the star, determine the system's 
{\it habitable zone}. For instance, 
planet's interior dynamics and atmospheric circulation generate a CO$_2$ cycle,
which subsequently results in greenhouse effect. The latter
helps the planet to maintain a uniform temperature. This process can, however, be disrupted
if the planet is too close or too far from the central star. In other words, 
the distance of the planet from the central star must be such that the 
amount of the radiation received by the planet allow liquid water to exist on 
its surface and in its atmosphere. The orbital elements of the planet, on the other hand,
have to ensure that this object will maintain a stable orbit at all times.

The width of a habitable zone and the location of its inner and outer 
boundaries vary with the luminosity of the central star and the planet's 
atmospheric circulation models \cite{Men03,Jones05,Jones06}. Conservatively, 
the inner edge of a habitable zone
can be considered as the distance closer than which water on the surface of 
the planet evaporates due to a runaway greenhouse effect. In the same
manner, the outer edge of the habitable zone is placed
at a distance where, in the absence of CO$_2$ clouds, runaway glaciation 
will freeze the water and creates permanent ice on the surface of the planet. 
Using these definitions of the inner and outer boundaries of a habitable
zone, \cite{Kasting93} have shown that a conservative range for the
habitable zone of the Sun would be
between 0.95 AU and 1.15 AU from this star (figure \ref{fig:23}). 
As noted by \cite{Jones05}, however, the outer edge of this region may 
extend to farther distances \cite{Forget97,Williams97,Mischna00} close to 4 AU.

Since the notion of habitability is based on life on Earth,
one can calculate the boundaries of the habitable zone of a star by
comparing its luminosity with that of the Sun. For a star with 
a surface temperature $T$ and radius $R$, the luminosity $L$ and
its brightness $F(r)$ at a distance $r$ are given by
\begin{eqnarray}
F(r)={1\over {4\pi}}L(R,T){r^{-2}}=\sigma{T^4}{R^2}{r^{-2}}\,,
\end{eqnarray}
\noindent
where $\sigma$ is the Boltzmann constant. Using equation (5) and the fact that Earth 
is in the habitable zone of the Sun,
the radial distances of the inner and outer edges of the habitable 
zone of a star can be obtained from
\begin{equation}
r\,=\, {\Bigl({{T}\over {T_{\rm S}}}\Bigr)^2}\,
{\Bigl({{R}\over {R_{\rm S}}}\Bigr)}\,{r_{E}}.
\end{equation}
\noindent
In this equation, $T_{\rm S}$ and $R_{\rm S}$ are the surface
temperature and radius of the Sun, respectively, and $r_E$ represents the
distance of Earth from the Sun (i.e., the inner and outer edges of Sun's
habitable zone).
Using equation (6), the habitable zone of a star can be defined as a region where  
an Earth-like planet receives the same amount of radiation as Earth
receives from the Sun, so that it can develop and maintain similar habitable 
conditions as those on Earth.

\subsection{Formation of Habitable Planets in S-Type Binaries}

\begin{figure}
\begin{center}
\includegraphics{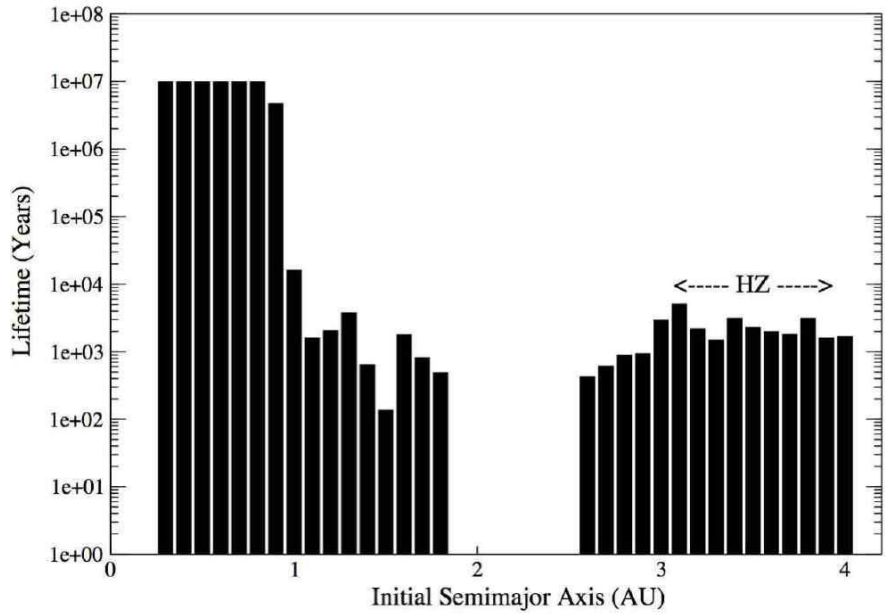}
\end{center}
\caption{Graph of the lifetime of an Earth-size object in
a circular orbit around the primary of $\gamma$ Cephei. The
habitable zone of the primary has been indicated by HZ. No 
planet was placed in the region between the aphelion and
perihelion distances of the giant planet of the system. As
shown here, only Earth-size planets close to the primary star
maintain their orbits for long times \cite{Hag06}.}
\label{fig:24}
\end{figure}

\begin{figure}
\begin{center}
\includegraphics{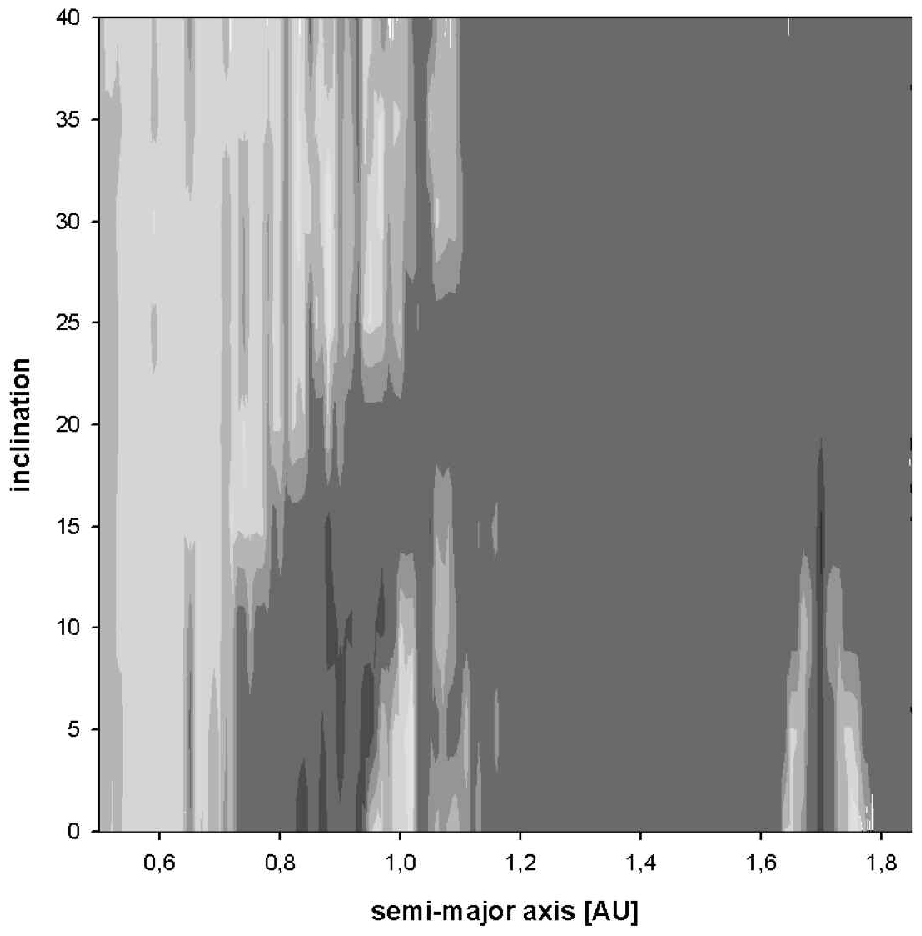}
\includegraphics{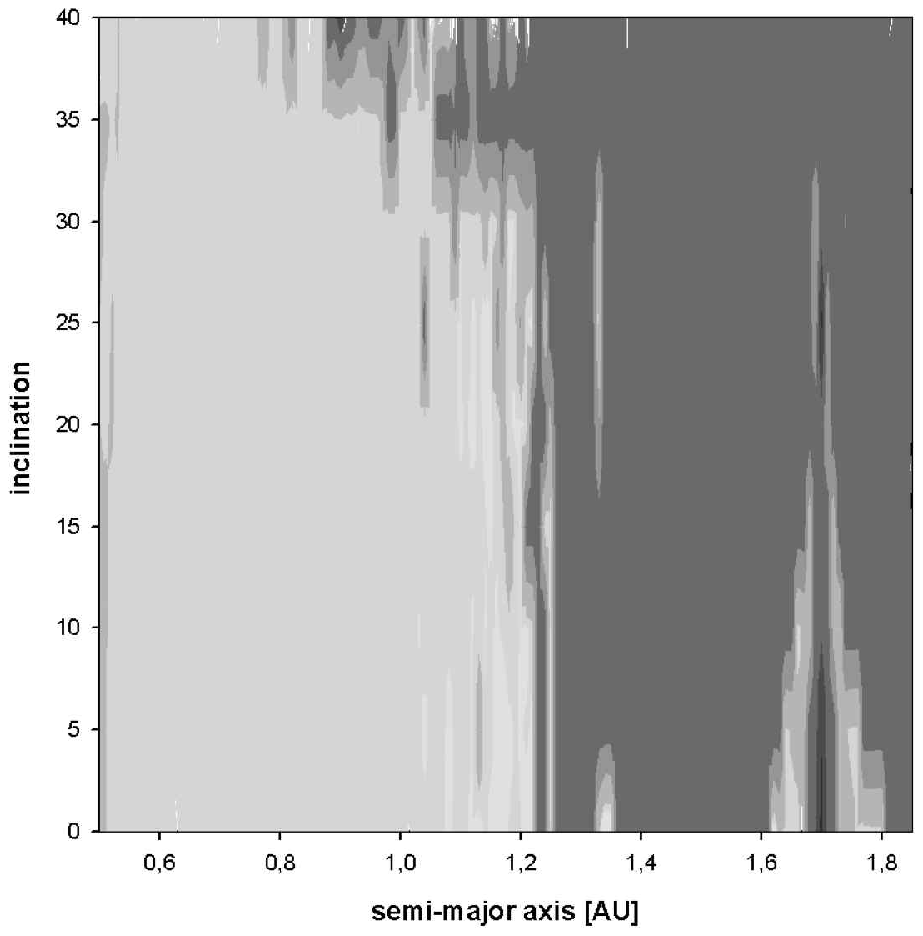}
\end{center}
\vskip -7.1in
\caption{Stability maps for a fictitious terrestrial planet in the 
$\gamma$ Cephei system. Dark regions represent chaotic zones 
and gray regions correspond to stability. The upper graph shows the stability
of a terrestrial planet  in a restricted four-body problem (R4BP) [i.e.\ $\gamma$ 
Cephei binary + detected planet + fictitious (massless) planet].  
The lower panel shows the results in a restricted three-body problem (R3BP) 
[i.e. primary + detected planet + fictitious planet] \cite{Pilat04}.}
\label{gamma}
\end{figure}

\begin{figure}
\vskip -2in
\begin{center}
\includegraphics{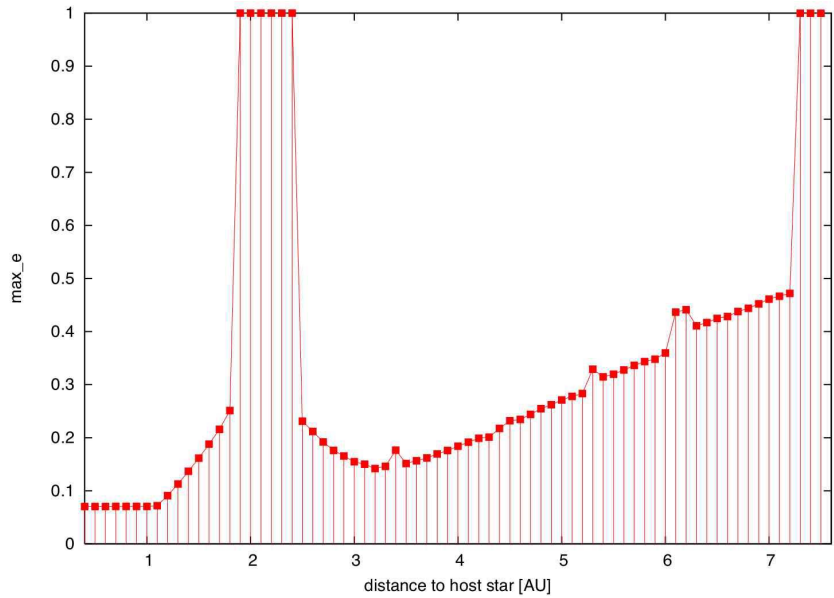}
\end{center}
\caption{Graph of the eccentricity of a massless terrestrial planet in a retrograde
orbit in the $\gamma$ Cephei system. Integrations were carried out for 1 Myr
within a restricted four-body system. The three peaks at 3.4 AU, 5.3 AU, and 6.1 AU
correspond to 2:1, 4:1, and 5:1 mean-motion resonances between the terrestrial
and the giant planets.}
 \label{TPio}
\end{figure}

\begin{figure}
\vskip -0.5in 
\begin{center}
\includegraphics{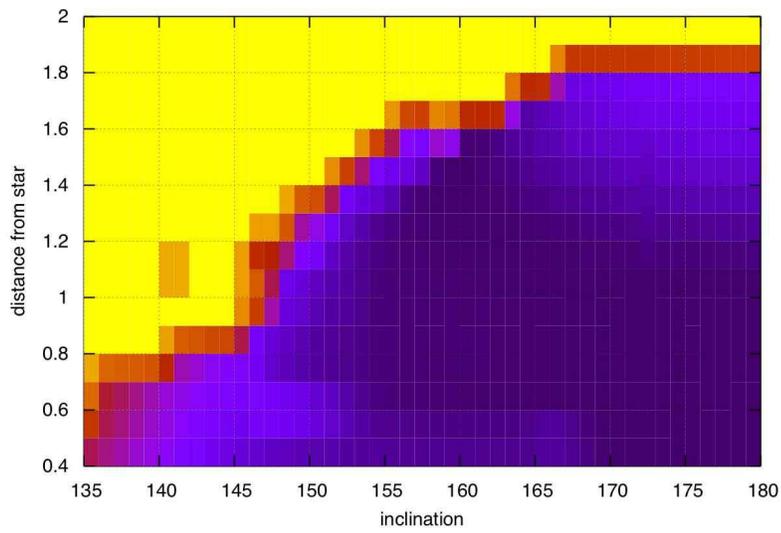}
\end{center}
\caption{Stability diagram for a retrograde TP-i orbit for different values of the orbital
inclination. The binary eccentricity is 0.35. Stability in shown in blue
whereas yellow corresponds to chaotic orbits.}
\label{TPi}
\end{figure}

\begin{figure}
\begin{center}
\includegraphics{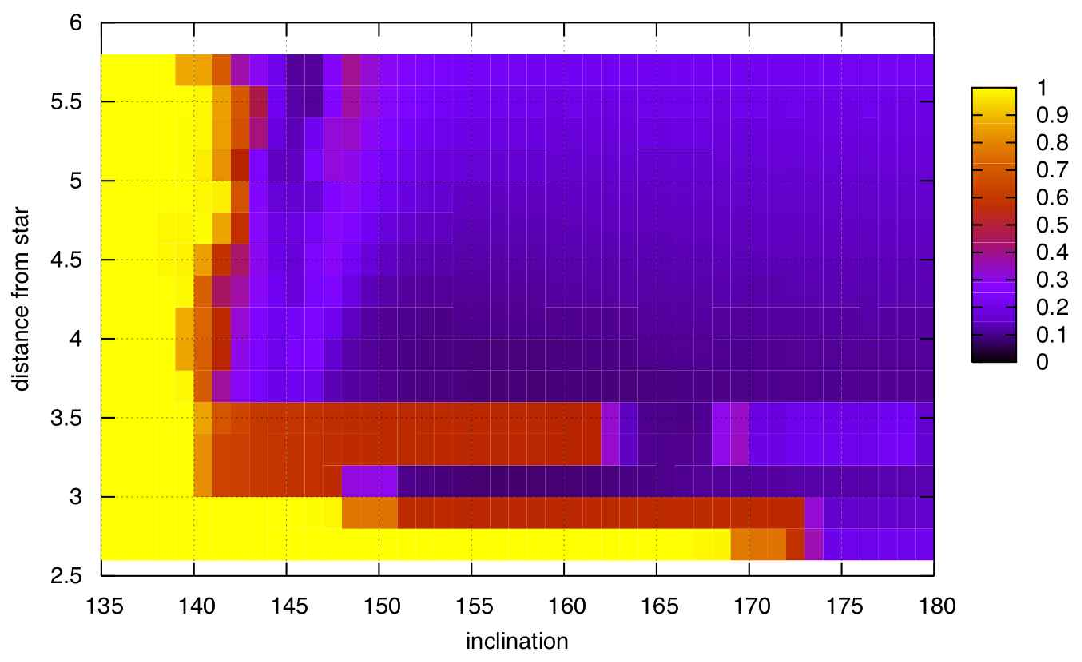}
\includegraphics{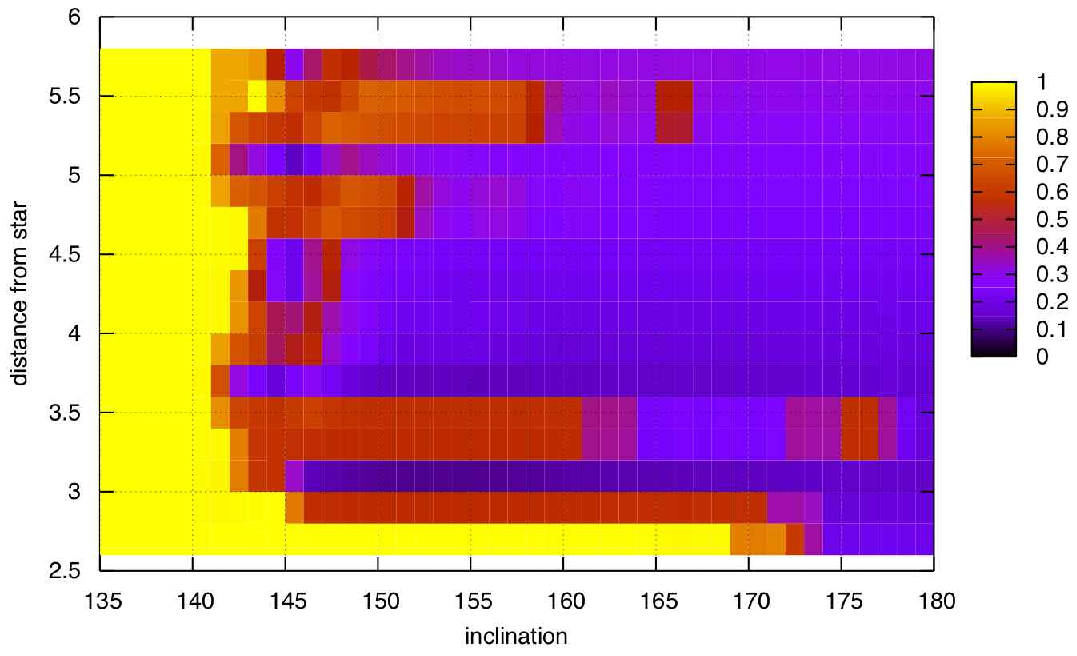}
\includegraphics{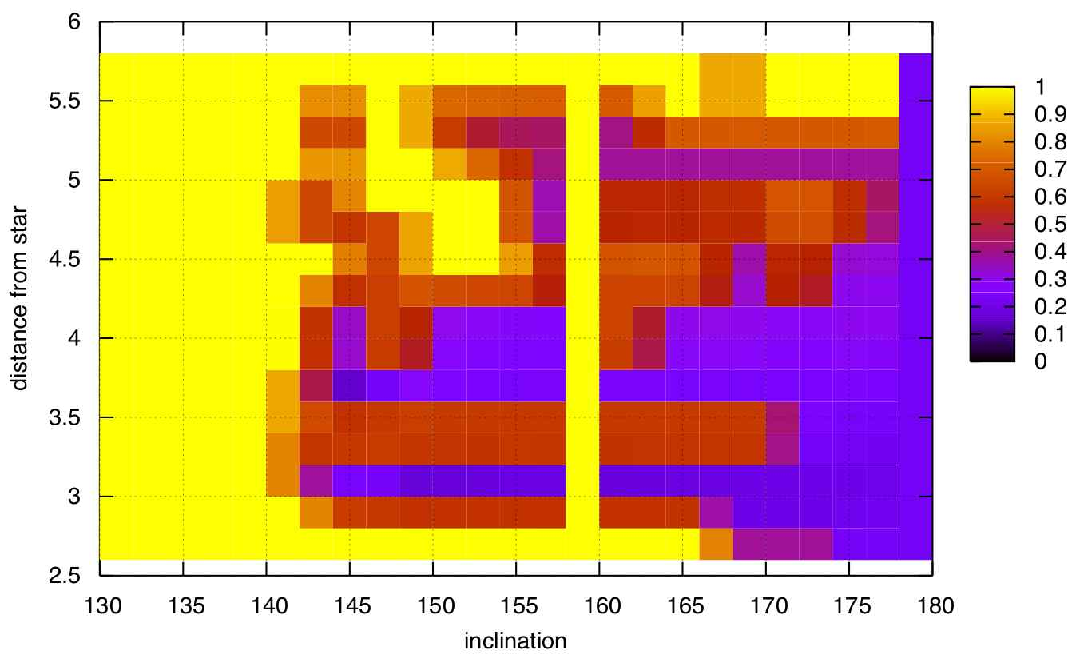}
\end{center}
\vskip -8.25in
\caption{Stability diagram for a retrograde TP-o orbit for different values of the orbital
inclination. The binary eccentricity is 0.25, 0.35, and 0.45 from top to bottom. 
Stability in shown in blue whereas yellow corresponds to chaotic orbits.
As shown here, instability extends to large distances as the binary eccentricity increases.}
\label{TPo}
\end{figure}

\begin{figure}
\vskip 1.5in
\begin{center}
\includegraphics{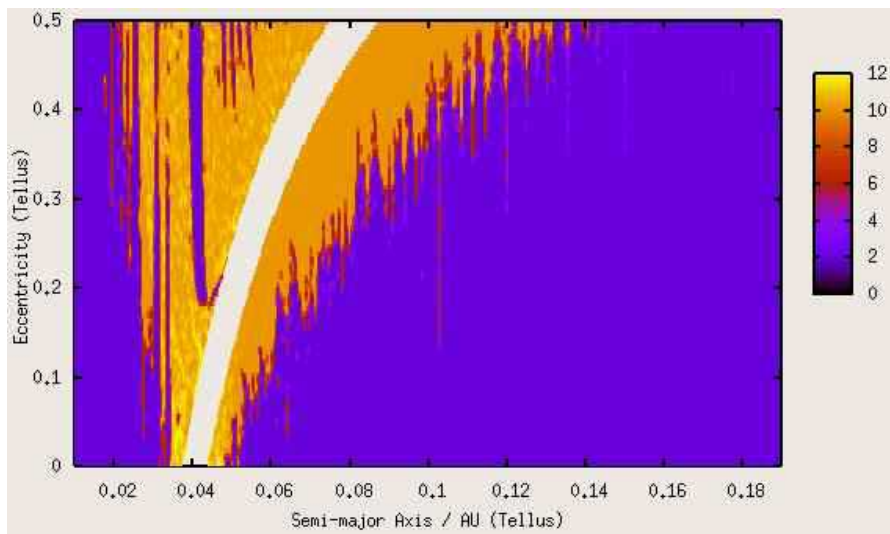}
\end{center}
\caption{Graph of the stability of an Earth-sized planet in a system consisting
of a Sun-like star and a transiting Jupiter-mass object in a 3 days orbit. Blue
shows stability whereas red corresponds to chaos. An island of stability
corresponding to the 1:1 mean-motion resonance is shown \cite{Hagh08,Capen09}. }
\label{megno}
\end{figure}

\begin{figure}
\vskip -2in
\begin{center}
\includegraphics{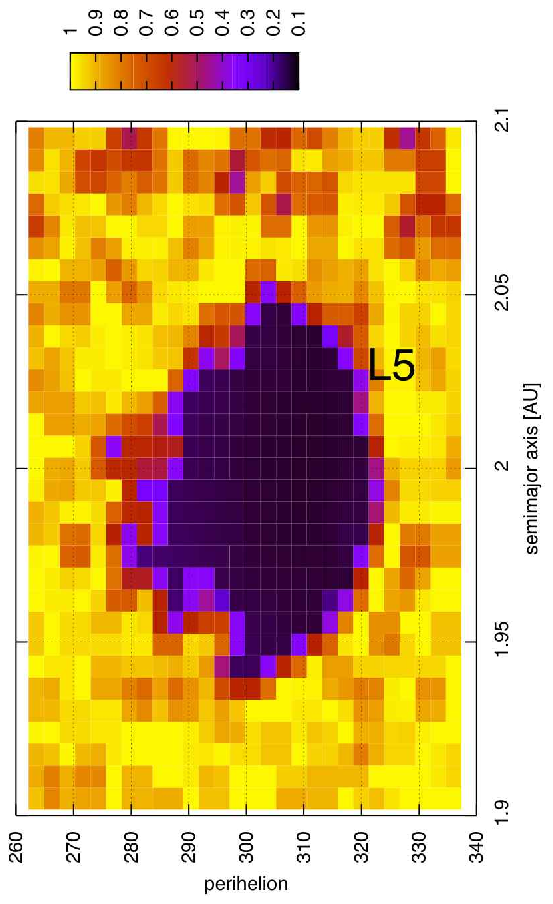}
\includegraphics{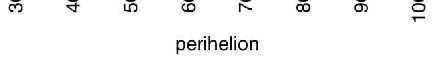}
\end{center}
\vskip -10.8in
\caption{Stability diagram for a TP-t orbit for the region around the Lagrange
points $L_4$ and $L_5$ of the giant planet in the $\gamma$ Cephei system.
The angular distance to the Lagrange point are shown on the horizontal axes
and the distance in AU is in the direction of the line connecting the  host-star
to the Lagrange point. The dark region shows the stable region.}
\label{l4-l5}
\end{figure}

\begin{figure}
\vskip -2in
\begin{center}
\includegraphics{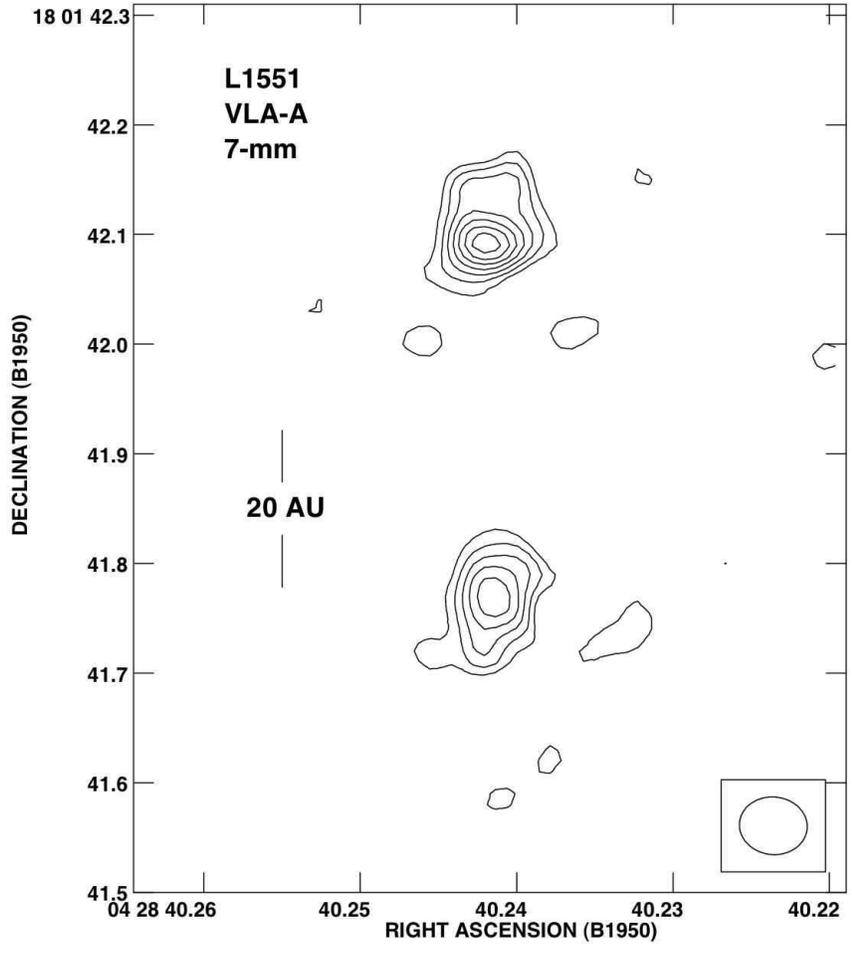}
\end{center}
\caption{Interferometric observation of the binary system L1551
\cite{Rodriguez98}. Two compact sources are evident in the map.
The separation of the binary is 45 AU and the disk around each 
core extends to approximately 10 AU. The mass of each disk is approximately
0.06 solar-masses.}
\label{fig:14}
\end{figure}

\begin{figure}
\begin{center}
\vskip 2in
\includegraphics{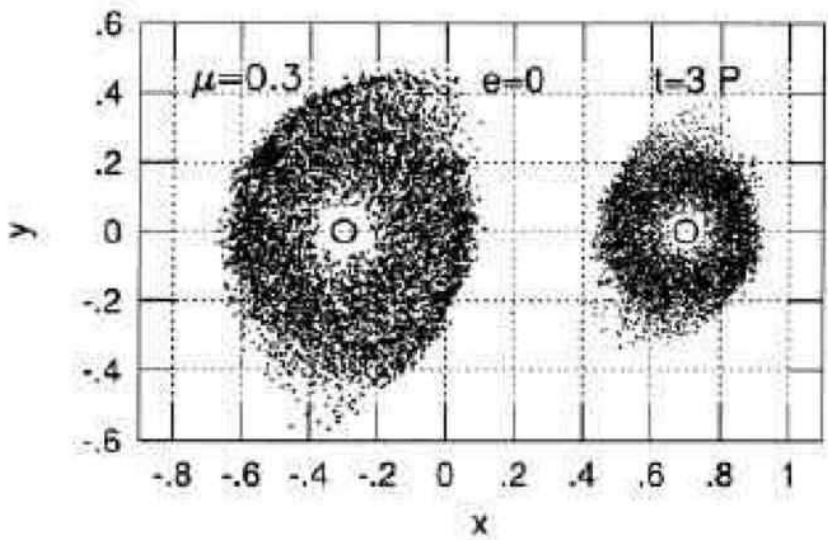}
\end{center}
\caption{Disk Truncation in and around binary systems \cite{Artymowicz94}
. The top graphs show circumstellar disk in a binary
with a mass-ratio of 0.3. Note the disk truncation when the eccentricity
of the binary is increased from 0 to 0.3. The bottom graphs show
similar effect in a circumbinary disk. The mass-ratio is 0.3 and the
binary eccentricity is 0.1. The numbers inside each graph represent 
the time in units of the binary period. The axes are in units
of the binary semimajor axis.}
\label{fig:13}
\end{figure}

\begin{figure}
\begin{center}
\includegraphics{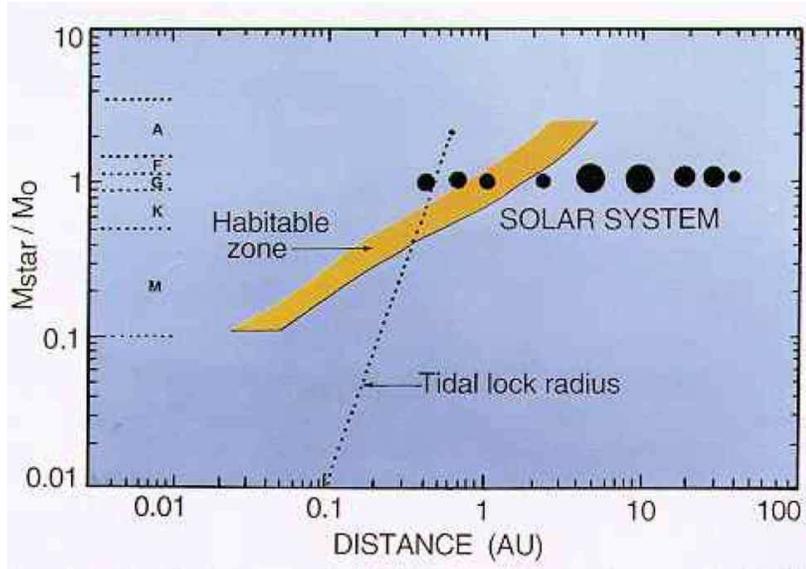}
\end{center}
\caption{Habitable zone \cite{Kasting93}.}
\label{fig:23}
\end{figure}

As mentioned earlier, a potentially habitable planet has to have a stable 
orbit in the habitable zone of its host star. Simulations of the stability of 
an Earth-size planet in the $\gamma$ Cephei system (figure \ref{fig:24}) indicate that 
in an S-type binary, the region of the stability of this object is
close to the primary, where the terrestrial planet is safe from the perturbations
of the giant planet and the secondary star. This implies that in order for the
primary to host a habitable planet, its habitable zone has to also fall within 
those distances where the orbit of an Earth-like planet is stable.
Within this framework, \cite{Hagh07} considered a binary star with a giant planet
in an S-type orbit and simulated the late stage of the formation of Earth-like planets
in the habitable zone of its primary star. In their simulations, these authors assumed that

\begin{figure}
\vskip 1.5in
\begin{center}
\includegraphics{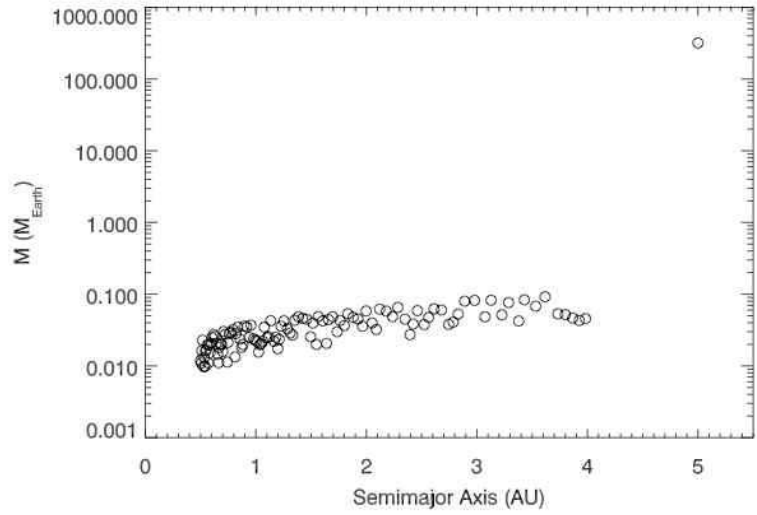}
\end{center}
\caption{Radial distribution of original protoplanetary objects \cite{Hagh07}.
\label{fig2}}
\end{figure}

\begin{itemize}

\item the primary star is Sun-like with a
habitable zone extending from 0.9 AU to 1.5 AU \cite{Kasting93}, 

\item a Jupiter-mass planet has already formed in a circular orbit at 5 AU from the
primary star,

\item the collisional growth of planetesimals has been efficient and has formed 
a disk of planetary embryos (e.g., via oligarchic growth \cite{Ida98}),

\item the water-mass fraction 
of embryos is similar to the current distribution of water in
primitive asteroids of the asteroid belt \cite{Abe00}. That is,
embryos inside 2 AU are dry, the ones
between 2 to 2.5 AU contain 1\% water, and those
beyond 2.5 AU have a water-mass fraction of 5\%
\cite{Raymond04,Raymond05a,Raymond05b,Raymond06a,Raymond06b}, and

\item the initial iron content for each embryo
is obtained by interpolating between the values of the
iron contents of the terrestrial planets
\cite{Lodders98,Raymond05a,Raymond05b}, with a dummy
value of 40\% in place of Mercury because of its anomalously high
iron content.

\end{itemize}
\noindent
The model  of \cite{Hagh07} also includes
a circumprimary disk of 115 Moon-to Mars-sized bodies, with masses ranging 
from 0.01 to 0.1 Earth-masses. These objects were
randomly distributed between 0.5 AU and 4 AU 
by 3 to 6 mutual Hill radii. The masses of embryos were increased with their
semimajor axes $(a)$ and the number of their mutual Hill radii
$(\Delta)$ as ${a^{3/4}}{\Delta^{3/2}}$ \cite{Raymond04}.
The surface density of the disk was assumed to vary as $r^{-1.5}$, where
$r$ is the radial distance from the primary star, and was normalized to
a density of 8.2 g/cm$^2$ at 1 AU. Figure \ref{fig2} shows the graph of 
one of such disks where the total mass is approximately 4 Earth-masses.

\begin{figure}
\vskip -1.5in
\begin{center}
\includegraphics{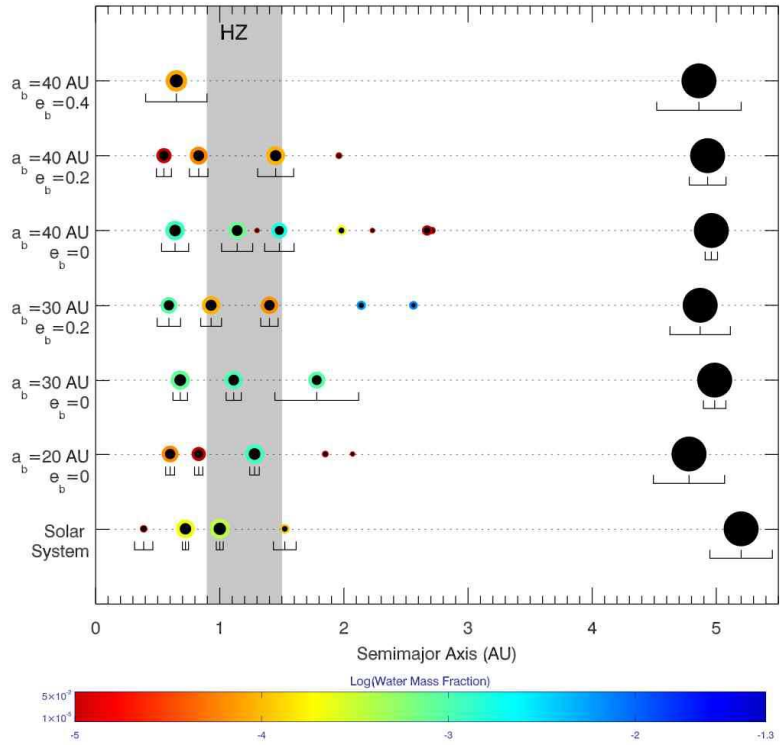}
\end{center}
\vskip -0.2in
\caption{Results of the simulations of habitable planet formation in a binary-planetary
system with ${\mu}=0.5$, for different values of the
eccentricity $(e_b)$ and semimajor axis $(a_b)$ of the stellar companion.
The inner planets of the solar system are shown for a comparison.
As seen from this figure, several Earth-like planets with substantial
amount of water are formed in the habitable zone of the star \cite{Hagh07}. 
\label{fig5}}
\end{figure}

The late stage of terrestrial planet formation \cite{Wetherill96}
was simulated by numerically integrating the orbits of the
planetary embryos for different values 
of the mass (0.5, 1.0, 1.5 solar-masses), semimajor axis (20, 30, 40 AU), 
and orbital eccentricity (0, 0.2, 0.4) of the secondary star.
The collisions among planetary embryos (which are the consequence of the
increase of their eccentricities due to their interactions with the secondary star
 \cite{Charnoz01} and the giant planet) were considered to be perfectly inelastic, 
 with no debris generated, 
 and no changes in the morphology and structures of the impacting bodies.
Similar to the current models of the formation of terrestrial planets
in our solar system, \cite{Hagh07} adopted the model of \cite{Morbidelli00} in which
water-rich bodies originating in the solar system's
asteroid belt were the primary source of Earth's water\footnote{It is 
important to emphasize that the
delivery of water to the inner part of the solar system might not have been
entirely due to the radial mixing of planetary embryos. Smaller
objects such as planetesimals in the outer region of the asteroid belt,
and comets originating in the outer solar system,
might have also contributed \cite{Raymond07b}.}. 
The delivery of water to a terrestrial planet was then facilitated by allowing transfer
of water from one embryo to another during their collision.
Figure \ref{fig5} shows the results of several of this simulations for a binary with a mass-ratio
of 0.5. The inner planets of the solar system are also shown for a comparison.
As shown here, several Earth-size planets, some with substantial amount of water,
are formed in the habitable zone of the primary star.

\begin{figure}
\vskip -1in
\begin{center}
\includegraphics{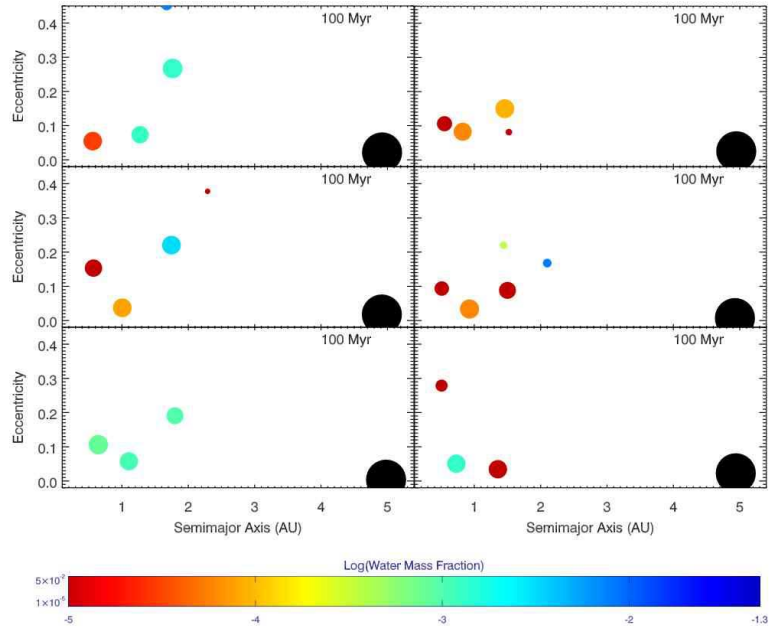}
\end{center}
\vskip -0.2in
\caption{The left column shows the
results of three simulations for different distribution of planetary embryos
in a binary with equal-mass Sun-like stars. The orbit
of the secondary star is circular with a radius of 30 AU. The right column shows 
the results of simulations for the same binary stars and similar distributions 
of planetary embryos where the secondary is in an orbit with a semimajor
axis of 40 AU and eccentricity of 0.2 \cite{Hagh07}.
\label{fig4}}
\end{figure}

An interesting result shown in figure \ref{fig5} is the 
relation between the orbital eccentricity of the stellar companion and the
water content of the final bodies. As shown here, in systems where the
secondary star has larger orbital eccentricity, the amount of water in
final planets is smaller. This can be seen more clearly in figure \ref{fig4},
where the final assembly and water contents of planets are shown
for a circular and an eccentric binary.
As shown here, for identical initial distributions of planetary embryos
(i.e., simulations on the same rows), the total water content of
the system on the left, where the secondary star is in a circular
orbit, is higher than that of the system on the right, where the
orbit of the secondary is eccentric. This can be attributed to the fact that
in an eccentric binary, because of the close approach of the secondary
star to the disk of planetary embryos, most of the water-carrying 
objects at the outer regions of the disk leave 
the system prior to the formation of terrestrial planets \cite{Artymowicz94,David03}. 
Simulations indicate that on average 90\% of
embryos in these systems were ejected during the integration
(i.e., their semimajor axes exceeded 100 AU) and among them, 60\%
collided with other protoplanetary bodies prior to their ejection
from the system. A small fraction of embryos $(\sim 5\%)$ also
collided with the primary or secondary star, or with the Jupiter-like
planet of the system \cite{Hagh07}.

\begin{figure}
\begin{center}
\includegraphics{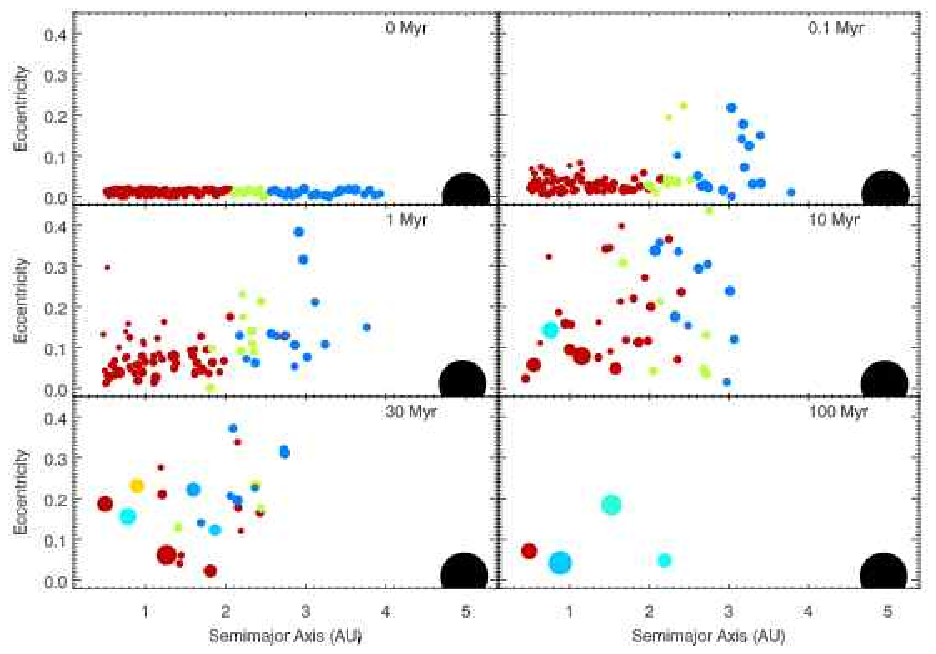}
\includegraphics{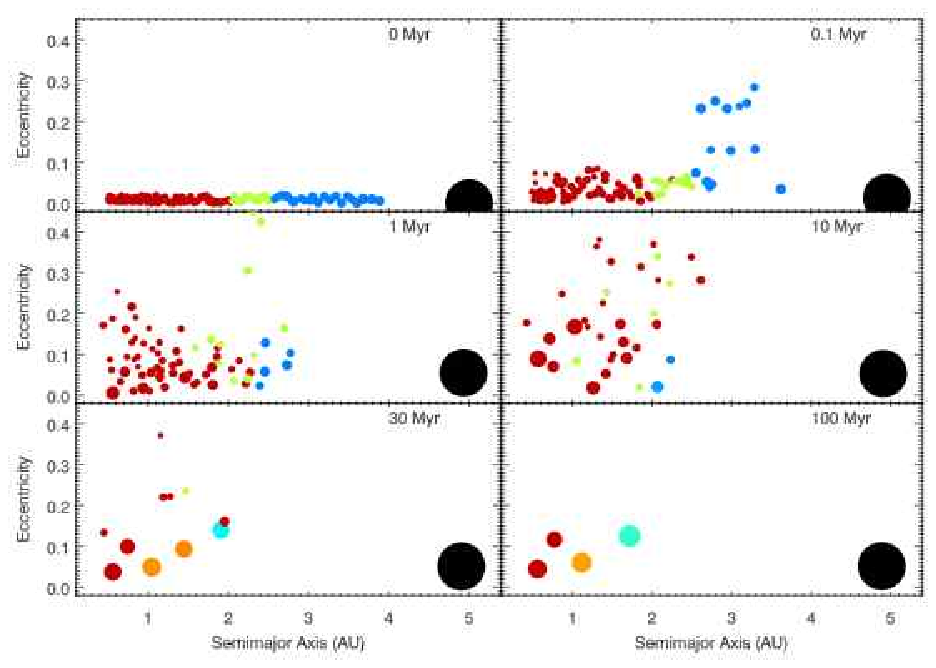}
\includegraphics{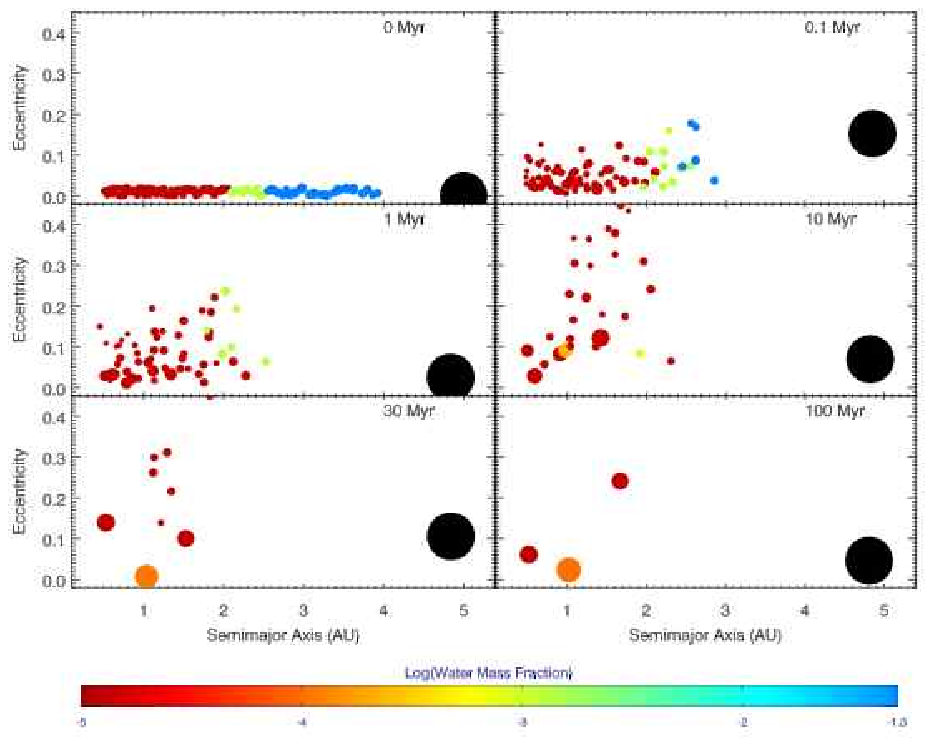}
\end{center}
\vskip -0.2in
\caption{Variation of water contents of the final planets with the eccentricity 
of the stellar companion. In these simulations, the primary star has a mass of 0.5
solar-masses, the semimajor axis of the binary is 30 AU,
and its eccentricity is equal to 0,0.2, and 0.4, from top to bottom \cite{Hagh07}.
\label{fig6}}
\end{figure}

\begin{figure}
\vskip 2in
\begin{center}
\includegraphics{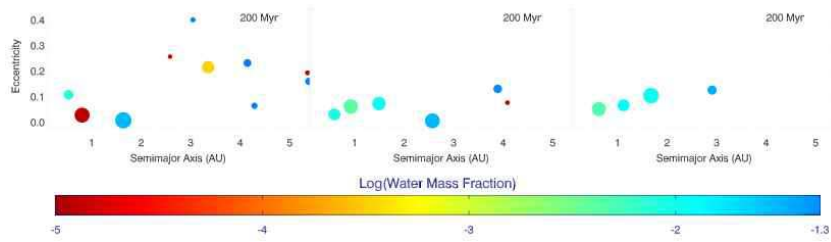}
\end{center}
\caption{Habitable planet formation in binary systems with no 
Jupiter-like planet. The stars of each binary are Sun-like
and their separations are 30 AU. The orbital eccentricity
of the secondary star is 0, 0.2, and 0.4, for the systems on 
the left, middle, and right, respectively. Note that compared with
previous simulations, the time of integration has to increase 
to 200 Myr in order to form comparable terrestrial-class planets \cite{Hagh07}.
\label{fig7}}
\end{figure}

In a binary-planetary system, the destabilizing effect of the secondary star is enhanced by 
the presence of the giant planet. Similar to our solar system, in these binaries, 
the Jovian-type planet perturbs the motion of embryos and enhances their radial mixing
and the rate of their collisions by
transferring angular momentum from the secondary star to these objects
\cite{Chambers02-II,Levison03,Raymond04,Raymond06a}.
Figure \ref{fig6} shows this in more details. The binary systems in these simulations
have  mass-ratios of 0.5, and their secondary stars are at 30 AU.
The binary eccentricity in these systems is equal
to 0, 0.2 and 0.4, from top to bottom. As shown here, as the
eccentricity  of the binary increases, the interaction of the secondary star with the
giant planet of the system becomes stronger (see the final eccentricity of
the giant planet), 
which causes closer approaches of this object to the disk of planetary embryos and
enhancing collisions and mixing among these bodies. The
eccentricities of embryos, at distances close to the outer edge of the
protoplanetary disk, rise to higher values until these bodies are
ejected from the system. In binaries with smaller perihelia,
the process of transferring angular momentum by means of the
giant planet is stronger and the ejection of protoplanets occurs
at earlier times. As a result, the total water budget of such systems is small.
A comparison between figure \ref{fig6} and figure \ref{fig7}, where simulations were
carried out for a binary without a giant planet, illustrates
the significance of the intermediate effect of the giant planet  
in a better way. As shown by figure \ref{fig7},
it is still possible to form terrestrial-class planets, with significant 
amount of water, in the habitable zone of the primary star. However, because 
of the lack of the transfer of angular momentum through the Jovian-type planet, 
the radial mixing of these objects is slower and terrestrial planet formation
takes longer. 

Another interesting result depicted by figure \ref{fig7} is the decrease in
the number of the final terrestrial planets and increase in their sizes and
accumulative water contents with increasing the eccentricity of
the secondary star. As shown here, from left panel to the right, as the
binary eccentricity increases, the close approach of the secondary star
to the protoplanetary disk increases the rate of the interaction of these objects
and enhances their collisions and radial mixing.
As a result, more of the water-carrying embryos
participate in the formation of the final terrestrial planets.
It is important to emphasize that
this process is efficient only in moderately eccentric binaries.
In binary systems with high eccentricities (small perihelia),
embryos may be ejected from the system \cite{David03}, and
terrestrial planet formation may become inefficient.

\begin{figure}
\begin{center}
\includegraphics{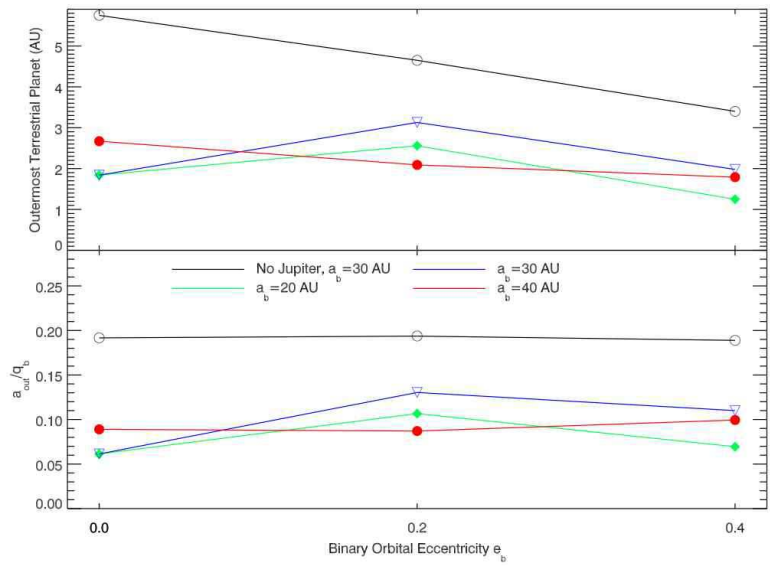}
\end{center}
\caption{Top panel: Semimajor axis of the outermost terrestrial planet.
Bottom panel: The ratio of the semimajor axis of this object 
to the perihelion distance of the binary. The secondary star is solar mass \cite{Hagh07}.
\label{fig8}}
\end{figure}

The results of the simulations without a giant planet imply a trend between
the location of the outer terrestrial planet and the perihelion of the binary.
In figure \ref{fig8} this has been shown for a set of different simulations. 
The top panel in this figure represents the semimajor axis of the outermost
terrestrial planet, $a_{\rm out}$, as a function of the binary
eccentricity, $e_b$. The bottom panel shows the ratio of
this quantity to the perihelion distance of the binary, $q_b$.
As shown here, terrestrial planet formation in binaries without a giant planet seems to
favor the region interior to approximately
 0.19 times the binary perihelion distance. This has also
been noted by \cite{Quintana07} (see their figure 9) in their simulations
of terrestrial planet formation in close binary star systems.
Given that the location of the inner edge of the habitable zone 
is at 0.9 AU, this trend implies that a binary perihelion distance of approximately
0.9/0.19 = 4.7 AU or larger may be necessary to allow habitable planet
formation. According to these results, habitable planet formation may not succeed
in binaries with Sun-like primaries that have stellar companions with
perihelion distances smaller than $\sim$5 AU. This, of course, is not a stringent 
condition, and is neither surprising since a moderately close
binary with a perihelion distance of $\sim$5 AU would be quite
eccentric, and as indicated by \cite{Hol99} the orbit
of a terrestrial-class object in the region around 1 AU from
the primary of such a system will be unstable [see equation (1)]\footnote{It is, 
also, important to note that, because the stellar luminosity, and therefore the 
location of the habitable zone, are sensitive to stellar mass
\cite{Kasting93,Raymond07b}, the minimum binary
separation necessary to ensure habitable planet formation
will vary significantly with the mass of the primary star.}.

In binary systems where a giant planet exist, figure \ref{fig8} indicates  that
terrestrial planets form closer-in. The ratio $a_{\rm {out}}/q_b$ in these
systems varies between approximately 0.06 and 0.13, depending on
the orbital separation of the two stars. The accretion process in
such systems is more complicated since the giant planet's eccentricity and
its ability to transfer angular momentum are largely regulated
by the binary companion.

\begin{figure}
\begin{center}
\includegraphics{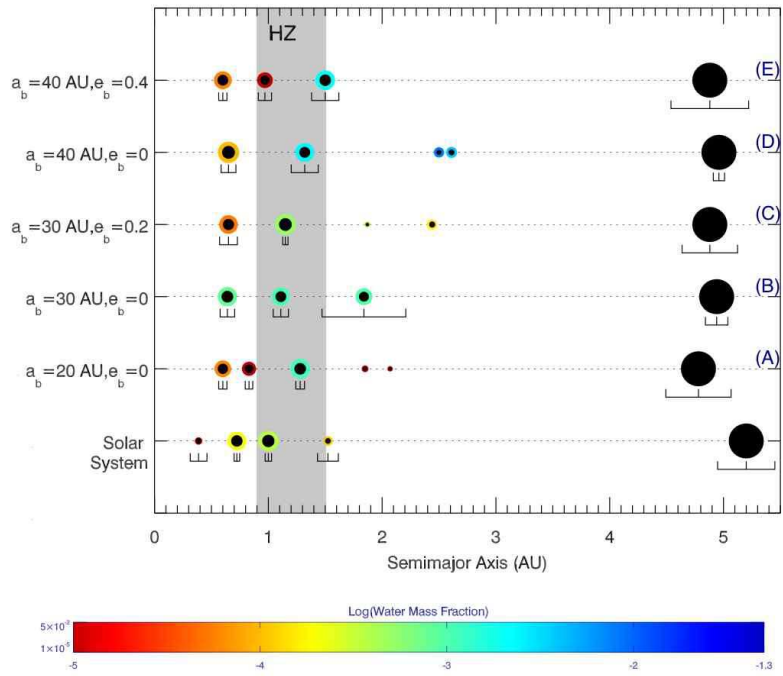}
\end{center}
\caption{Variation of number and water content of Earth-like objects with the
perihelion of the secondary star.
The mass of the secondary star in all simulations is 1 solar-mass \cite{Hagh07}.
\label{fig9}}
\end{figure}

\begin{figure}
\begin{center}
\includegraphics{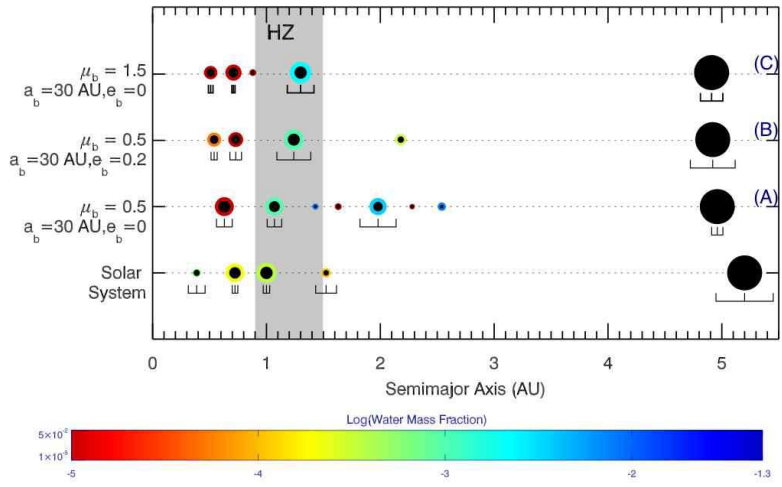}
\end{center}
\caption{Variation of number and water content of Earth-like objects with the
mass-ratio of the binary \cite{Hagh07}.
\label{fig10}}
\end{figure}

As shown by \cite{Hagh07}, despite the stochasticity of the simulations,
and the large size of the parameter-space, many simulations
resulted in the formation of Earth-like planet, with substantial amount of water,
in the habitable zone of the primary star. Figures \ref{fig9} and \ref{fig10} 
show the result. The orbital elements of the final objects are given in Table \ref{tab:4}.
It is necessary to mention that because in these simulations,
all collisions have been considered to be perfectly inelastic (i.e.,
the water contents of the resulted planets would be equal to the sum of the 
water contents of the impacting bodies, and the loss of water due to the impact 
and the motion of the ground of an impacted body  \cite{Genda05,Canup06} has
been ignored), the numbers given in Table \ref{tab:4} set an upper limit
for the water budget of final planets. The total water budget of these objects
may in fact be 5-10 times smaller than those reported here \cite{Raymond04}.

\begin{table}
\centering
\caption{(Sun's habitable zone: 0.9 - 1.50 AU)}
\label{tab:4} 
\begin{tabular}{ccccccccccccc}
\hline\noalign{\smallskip}
Simulation &  & & ${m_p}\,(M_\oplus)$ & & & $a_p$ (AU) & & & $e_p$ & & & Water Fraction  \\
\noalign{\smallskip}\hline\noalign{\smallskip}
9-A    & & & 0.95    & & & 1.28     & & & 0.03    & & & 0.00421       \\
9-B    & & & 0.75    & & & 1.11     & & & 0.06    & & & 0.00415       \\
9-C    & & & 1.17    & & & 1.16     & & & 0.03    & & & 0.00164       \\
9-D    & & & 0.86    & & & 1.33     & & & 0.09    & & & 0.01070       \\
9-E    & & & 0.95    & & & 1.50     & & & 0.08    & & & 0.00868       \\
10-A   & &  & 0.74    & & & 1.07     & & & 0.06    & &  & 0.00349       \\
10-B   & & & 0.99    & & & 1.26     & & & 0.12    & & & 0.00366       \\
10-C   & & & 1.23    & & &1.30      & & & 0.09    & & &0.00103       \\
\noalign{\smallskip}\hline
\end{tabular}
\end{table}

\begin{figure}
\vskip 1in
\begin{center}
\includegraphics{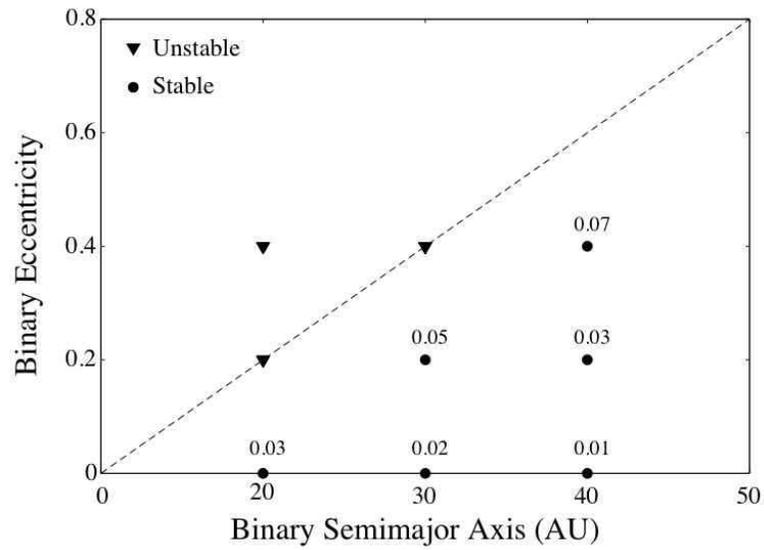}
\end{center}
\caption{Habitable planet formation in the $({e_b},{a_b})$ space
of an equal-mass binary-planetary system.
Circles correspond to binaries in which habitable planets are formed. Triangles
represent systems in which the giant planet is unstable.
The number associated with each circle represents the average
eccentricity of the giant planet of the system during the simulation \cite{Hagh07}.
\label{fig11}}
\end{figure}

A study of the systems of figures \ref{fig9} and \ref{fig10} indicates that these binaries 
have relatively large perihelia. Figure \ref{fig11} shows this for simulations in a
binary with a mass-ratio of 0.5 in terms of the semimajor axis and eccentricity of 
the stellar companion. The circles in this figure represent those systems in which the 
giant planet maintained a stable orbit and also simulations resulted in the formation of 
habitable bodies. The number associated with each circle corresponds to
the average eccentricity of the giant planet during the simulation.
The triangles correspond to systems in which the giant planet became unstable.
Given that at the beginning of each simulation, the orbit of the giant planet 
was circular, a non-zero value for its average eccentricity is indicative
of its interaction with the secondary star. The fact that Earth-like objects were 
formed in systems where the average eccentricity of the giant planet is small implies that
this interaction has been weak. In other words, binaries with moderate to large 
perihelia and with giant planets on low eccentricity orbits are most favorable
for habitable planet formation. Similar to the formation of habitable planets around 
single stars, where giant planets, in general, play destructive roles, a strong interaction
between the secondary star and the giant planet in a binary-planetary system
(i.e., a small binary perihelion) increases the orbital
eccentricity of this object, and results in the removal of the terrestrial
planet-forming materials from the system. For more details we refer the reader
to \cite{Hagh07}.

\input{Hagh-referenc}

\printindex

\clearpage

\clearpage

\clearpage

\clearpage

\clearpage

\clearpage

\clearpage

\clearpage

\clearpage

\clearpage

\clearpage

\clearpage

\clearpage

\clearpage

\clearpage

\clearpage

\clearpage

\clearpage

\clearpage

\clearpage

\clearpage

\clearpage

\clearpage

\clearpage

\clearpage

\clearpage

\clearpage

\clearpage

\clearpage

\end{document}

%% file: Hagh-referenc.tex
%
%

%
%